\documentclass[useAMS,usenatbib]{mn2e}

\usepackage{graphicx}  
\usepackage{ amssymb }
\usepackage{journals}

\newcommand{\teff}{$T_{\rm eff}$} 
\newcommand{\logg}{$\log g$} 
\newcommand{\kms}{km s$^{-1}$}
\newcommand{\vt}{$\xi_t$} 
\newcommand{\fei}{Fe\,{\sc i}}
\newcommand{\feii}{Fe\,{\sc ii}}
\newcommand{\tii}{Ti\,{\sc i}}
\newcommand{\tiii}{Ti\,{\sc ii}}
\newcommand{\cri}{Cr\,{\sc i}}
\newcommand{\crii}{Cr\,{\sc ii}}

\newcommand\simgt{\lower.3ex\hbox{\gtsima}}
\newcommand\mlogg{\rm log g}
\newcommand\Pd{\partial}
\newcommand\ep{\rm{log}\epsilon}

\title[Precision Abundance Measurements in NGC 6752]{High Precision
Differential Abundance Measurements in Globular Clusters: Chemical
Inhomogeneities in NGC~6752\thanks{Based on observations collected at the
European Southern Observatory, Chile (ESO Programmes 67.D-0145 and
65.L-0165A).}} 
\author[D.\ Yong et al.]
{David Yong$^{1}$\thanks{E-mail: yong@mso.anu.edu.au}, 
Jorge Mel{\' e}ndez$^2$, 
Frank Grundahl$^3$, 
Ian U.\ Roederer$^4$, \newauthor 
John E.\ Norris$^1$, 
A.\ P.\ Milone$^1$, 
A.\ F.\ Marino$^1$,  
P.\ Coelho$^5$, \newauthor 
Barbara E.\ McArthur$^6$, 
K.\ Lind$^7$, 
R.\ Collet$^1$ and 
Martin Asplund$^1$. \\ 
$^{1}$Research School of Astronomy and Astrophysics, Australian
National University, Canberra, ACT 2611, Australia\\ 
$^{2}$Departamento de Astronomia do IAG/USP, Universidade
de Sao Paulo, Rua do Matao 1226, Sao Paulo, 05508-900, SP, Brasil\\
$^{3}$Stellar Astrophysics Centre, Department of Physics and 
Astronomy, Aarhus University, Ny Munkegade 120, DK-8000 Aarhus C, Denmark\\
$^{4}$Carnegie Observatories, 813 Santa Barbara Street, Pasadena, CA
91101, USA \\
$^{5}$N\'ucleo de Astrof{\' i}sica Te{\' o}rica, Universidade Cruzeiro do Sul,
R. Galv{\~ a}o Bueno 868, Liberdade 01506-000, S{\~ a}o Paulo, Brazil\\ 
$^{6}$McDonald Observatory, University of Texas, Austin, TX 78712, USA\\
$^{7}$University of Cambridge, Madingley Road, Cambridge, CB3 0HA, UK 
}
\begin{document}

\pagerange{\pageref{firstpage}--\pageref{lastpage}} \pubyear{2013}

\maketitle

\label{firstpage}

\begin{abstract}
We report on a strictly differential line-by-line analysis of high quality UVES
spectra of bright giants in the metal-poor globular cluster NGC 6752. We
achieved high precision differential chemical abundance measurements for Fe,
Na, Si, Ca, Ti, Cr, Ni, Zn, Y, Zr, Ba, La, Ce, Pr, Nd, Sm, Eu and Dy with
uncertainties as low as $\sim$0.01~dex ($\sim$2\%). We obtained the following
main results.  (1) The observed abundance dispersions are a factor of $\sim$2
larger than the average measurement uncertainty. (2) There are positive
correlations, of high statistical significance, between all elements and Na.
(3) For any pair of elements, there are positive correlations of high
statistical significance, although the amplitudes of the abundance variations
are small.  Removing abundance trends with effective temperature and/or using a
different pair of reference stars does not alter these results.  These
abundance variations and correlations may reflect a combination of ($a$) He
abundance variations and $(b)$ inhomogeneous chemical evolution in the pre- or
proto-cluster environment. Regarding the former, the current constraints on
$\Delta Y$ from photometry likely preclude He as being the sole explanation.
Regarding the latter, the nucleosynthetic source(s) must have synthesised Na,
$\alpha$, Fe-peak and neutron-capture elements and in constant amounts for
species heavier than Si; no individual object can achieve such nucleosynthesis.
We speculate that other, if not all, globular clusters may exhibit comparable
abundance variations and correlations to NGC 6752 if subjected to a similarly
precise analysis. 
\end{abstract}


\section{Introduction}

Understanding the origin of the star-to-star abundance variations of the light
elements in globular clusters is one of the major challenges confronting
stellar evolution, stellar nucleosynthesis and chemical evolution. Arguably the
first evidence for chemical abundance inhomogeneity in a globular cluster was
the discovery of a CN strong star in M13 by \citet{popper47}. A large number of
subsequent studies have confirmed the star-to-star variation in the strength of
the CN molecular bands in a given globular cluster, and these results have been
extended to star-to-star abundance variations for the light elements -- Li, C,
N, O, F, Na, Mg and Al (e.g., see reviews by \citealt{smith87},
\citealt{kraft94} and \citealt{gratton04,gratton12}). In light of the discovery
of abundance variations in unevolved stars (e.g.,
\citealt{cannon98,gratton01,ramirez02,ramirez03}), the consensus view is that
these light element abundance variations are attributed to a proto-cluster
environment in which the gas was of an inhomogeneous composition. The
interstellar medium from which some of the stars formed included material
processed through hydrogen-burning at high temperatures. The source of that
material and the nature of the nucleosynthesis, however, remain highly
contentious with intermediate-mass asymptotic giant branch stars, fast rotating
massive stars and massive binaries being the leading candidates (e.g.,
\citealt{fenner04}, \citealt{ventura05}, \citealt{decressin07},
\citealt{demink09}, \citealt{marcolini09}). 

Recent discoveries of complex structure in colour-magnitude diagrams reveal
that most, if not all, globular clusters host multiple populations; the
evidence consists of multiple main sequences, subgiant branches, red giant
branches and/or horizontal branches in Galactic (e.g., see \citealt{piotto09}
for a review) and also extragalactic globular clusters (e.g.,
\citealt{mackey07,milone09}). When using appropriate photometric filters, all
globular clusters show well-defined sequences with distinct chemical abundance
patterns \citep{milone12}. These multiple populations can be best explained by
different ages and/or chemical compositions. The sequence of events leading to
the formation of multiple population globular clusters is not well understood
(e.g., \citealt{dercole08,bekki11,conroy11}). 

Although the census and characterization of the Galactic globular clusters
remains incomplete, they may be placed into three general
categories\footnote{There are subtle, and not so subtle, differences within a
given category.}: 
($i$) those that exhibit only light element abundance variations, which include
NGC 6397, NGC 6752 and 47~Tuc (e.g.,
\citealt{gratton01,yong05,dorazi10,lind11,campbell13}), 
($ii$) those that exhibit light element abundance variations and
neutron-capture element abundance dispersions such as M15 (e.g.,
\citealt{sneden97,sneden00,sobeck11}) and 
($iii$) those that exhibit light element abundance variations as well as
significant abundance dispersions for Fe-peak
elements\footnote{\citet{saviane12} have identified a metallicity dispersion in
NGC 5824. To our knowledge, there are no published studies of the light element
abundances based on high-resolution spectroscopy, so we cannot yet place this
globular cluster in category ($iii$).} such as $\omega$ Cen, M22, M54, NGC
1851, NGC 3201 and Terzan 5 (e.g.,
\citealt{norris95,yong081851,marino09,marino11,carretta10,johnson10,villanova10,carretta11,origlia11,roederer11,alvesbrito12,simmerer13}).
At this stage, we do not attempt to classify a particularly unusual system like
NGC 2419 \citep{cohen10,cohen11,cohen12,mucciarelli12}. 

Given the surprisingly large star-to-star variations in element abundance
ratios in a given cluster, how chemically homogeneous are the ``well-behaved''
elements in the "normal" globular clusters (i.e., clusters in category ($i$)
above)? The answer to this question has important consequences for testing
model predictions, setting constraints on the polluters and understanding the
origin and evolution of globular clusters. 

\citet{sneden05} considered the issue of cluster abundance accuracy limits and
selected the [Ni/Fe] ratio as an example. This pair of elements was chosen as
they present numerous spectral lines in the ``uncomplicated yellow-red region''
of the spectrum and share ``common nucleosynthetic origins in supernovae''.
\citet{sneden05} noted that the dispersion in the [Ni/Fe] ratio in a cluster
was $\sim$0.06~dex and appeared to show ``little apparent trend as a function
of the number of stars observed in a survey or of year of publication''. There
are two possible reasons for the apparent limit in the $\sigma$[Ni/Fe] ratio.
Perhaps clusters possess a single [Ni/Fe] ratio and the dispersion reflects the
measurement uncertainties. Alternatively, globular clusters are chemically
homogeneous in the [Ni/Fe] ratio at the $\sim$0.06~dex level.  Bearing in mind
this apparent limit in the [Ni/Fe] dispersion, in order to answer the question
posed above, we require the highest possible precision when measuring chemical
abundances. 

A number of recent studies have achieved precision in chemical abundance
measurements as low as 0.01~dex (e.g., \citealt{melendez09,melendez12},
\citealt{alvesbrito10}, \citealt{nissen10,nissen11},
\citealt{ramirez10,ramirez12}). These results were obtained by using ($i$) high
quality spectra (R $\ge$ 60,000 and signal-to-noise ratios S/N $\ge$ 200 per
pixel), ($ii$) a strictly differential line-by-line analysis and ($iii$) a
well-chosen sample of stars covering a small range in stellar parameters
(effective temperature, surface gravity, metallicity). Application of similar
analysis techniques to high quality spectra of stars in globular clusters
offers the hope that high precision chemical abundance measurements (at the
$\sim$0.01 dex level) can also be obtained. To our knowledge, the highest
precision chemical abundance measurements in globular clusters to date are at
the $\sim$0.04 dex level include \citet{yong05}, \citet{gratton05},
\citet{carretta09} and \citet{melendez09m71}.  The aim of the present paper is
to achieve high precision abundance measurements in the globular cluster NGC
6752 and to use these data to study the chemical enrichment history of this
cluster.

\section{OBSERVATIONS AND ANALYSIS}
\label{sec:obs}

\subsection{Target Selection and Spectroscopic Observations} 

The targets for this study were taken from the $uvby$ photometry by
\citet{grundahl99}. The sample consists of 17 stars located near the tip of the
red giant branch (hereafter RGB tip stars) and 21 stars located at the bump in
the luminosity function along the RGB (hereafter RGB bump stars). The list of
targets can be found in Table \ref{tab:param}. Observations were performed
using the Ultraviolet and Visual Echelle Spectrograph (UVES;
\citealt{dekker00}) on the 8.2m Kueyen (VLT/UT2) telescope at Cerro Paranal,
Chile. The RGB tip stars were observed at a resolving power of R = 110,000 and
S/N $\ge$ 150 per pixel near 5140\AA\ while the RGB bump stars were observed at
R = 60,000 and S/N $\ge$ 100 per pixel near 5140\AA. Analyses of these spectra
have been reported in \citet{grundahl02} and \citet{yong03,yong05,yong08nh}.
The location of the program stars in a colour-magnitude diagram can be found in
Figure 1 in \citet{yong03}. 

\begin{table*}
 \centering
 \begin{minipage}{140mm}
  \caption{Program Stars and Stellar Parameters as Defined in Section 2.3.}
  \label{tab:param} 
  \begin{tabular}{@{}llccccccc@{}}
  \hline
	Name1\footnote{PD1 and PD2 are from \citet{penny86} and BXXXX names are
from \citet{buonanno86}.}  & 
	Name2 & 
	RA2000 & 
	DE2000 & 
	$V$ & 
	\teff\footnote{These stellar parameters are for the so-called ``reference
star'' values (see Section 2.3 for details).}  & 
	\logg$^b$ & 
	\vt$^b$ & 
	[Fe/H]$^b$ \\ 
	 & 
	 & 
	 & 
	 & 
	 & 
	(K) & 
	(cm s$^{-2}$) & 
	(\kms) & 
	 \\
	(1) & 
	(2) &
	(3) &
	(4) & 
	(5) & 
	(6) & 
	(7) & 
	(8) & 
	(9) \\ 
  \hline
   PD1 &  NGC6752-mg0 & 19:10:58 & $-$59:58:07 & 10.70 & 3928 & 0.26 & 2.20 & $-$1.67 \\
 B1630 &  NGC6752-mg1 & 19:11:11 & $-$59:59:51 & 10.73 & 3900 & 0.24 & 2.25 & $-$1.70 \\
 B3589 &  NGC6752-mg2 & 19:10:32 & $-$59:57:01 & 10.94 & 3894 & 0.33 & 2.07 & $-$1.66 \\
 B1416 &  NGC6752-mg3 & 19:11:17 & $-$60:03:10 & 10.99 & 4050 & 0.50 & 1.88 & $-$1.66 \\
\ldots &  NGC6752-mg4 & 19:10:43 & $-$59:59:54 & 11.02 & 4065 & 0.53 & 1.86 & $-$1.65 \\
   PD2 &  NGC6752-mg5 & 19:10:49 & $-$59:59:34 & 11.03 & 4100 & 0.56 & 1.90 & $-$1.65 \\
 B2113 &  NGC6752-mg6 & 19:11:03 & $-$60:01:43 & 11.22 & 4154 & 0.68 & 1.85 & $-$1.62 \\
\ldots &  NGC6752-mg8 & 19:10:38 & $-$60:04:10 & 11.47 & 4250 & 0.80 & 1.71 & $-$1.69 \\
 B3169 &  NGC6752-mg9 & 19:10:40 & $-$59:58:14 & 11.52 & 4288 & 0.91 & 1.72 & $-$1.66 \\
 B2575 & NGC6752-mg10 & 19:10:54 & $-$59:57:14 & 11.54 & 4264 & 0.90 & 1.66 & $-$1.67 \\
\ldots & NGC6752-mg12 & 19:10:58 & $-$59:57:04 & 11.59 & 4286 & 0.94 & 1.73 & $-$1.68 \\
 B2196 & NGC6752-mg15 & 19:11:01 & $-$59:57:18 & 11.68 & 4354 & 1.02 & 1.74 & $-$1.64 \\
 B1518 & NGC6752-mg18 & 19:11:15 & $-$60:00:29 & 11.83 & 4398 & 1.11 & 1.68 & $-$1.64 \\
 B3805 & NGC6752-mg21 & 19:10:28 & $-$59:59:49 & 11.99 & 4429 & 1.20 & 1.68 & $-$1.65 \\
 B2580 & NGC6752-mg22 & 19:10:54 & $-$60:02:05 & 11.99 & 4436 & 1.20 & 1.71 & $-$1.65 \\
 B1285 & NGC6752-mg24 & 19:11:19 & $-$60:00:31 & 12.15 & 4511 & 1.31 & 1.69 & $-$1.67 \\
 B2892 & NGC6752-mg25 & 19:10:46 & $-$59:56:22 & 12.23 & 4489 & 1.33 & 1.70 & $-$1.67 \\
\ldots &    NGC6752-0 & 19:11:03 & $-$59:59:32 & 13.03 & 4699 & 1.83 & 1.43 & $-$1.66 \\
 B2882 &    NGC6752-1 & 19:10:47 & $-$60:00:43 & 13.27 & 4749 & 1.95 & 1.37 & $-$1.63 \\
 B1635 &    NGC6752-2 & 19:11:11 & $-$60:00:17 & 13.30 & 4779 & 1.98 & 1.37 & $-$1.63 \\
 B2271 &    NGC6752-3 & 19:11:00 & $-$59:56:40 & 13.41 & 4796 & 2.03 & 1.38 & $-$1.69 \\
  B611 &    NGC6752-4 & 19:11:33 & $-$60:00:02 & 13.42 & 4806 & 2.04 & 1.38 & $-$1.65 \\
 B3490 &    NGC6752-6 & 19:10:34 & $-$59:59:55 & 13.47 & 4804 & 2.06 & 1.33 & $-$1.64 \\
 B2438 &    NGC6752-7 & 19:10:57 & $-$60:00:41 & 13.53 & 4829 & 2.10 & 1.32 & $-$1.86\footnote{We exclude this star from the subsequent differential analysis due to its discrepant metallicity.} \\
 B3103 &    NGC6752-8 & 19:10:45 & $-$59:58:18 & 13.56 & 4910 & 2.15 & 1.33 & $-$1.69 \\
 B3880 &    NGC6752-9 & 19:10:26 & $-$59:59:05 & 13.57 & 4824 & 2.11 & 1.41 & $-$1.70 \\
 B1330 &   NGC6752-10 & 19:11:18 & $-$59:59:42 & 13.60 & 4836 & 2.13 & 1.37 & $-$1.65 \\
 B2728 &   NGC6752-11 & 19:10:50 & $-$60:02:25 & 13.62 & 4829 & 2.13 & 1.34 & $-$1.68 \\
 B4216 &   NGC6752-12 & 19:10:20 & $-$60:00:30 & 13.64 & 4841 & 2.15 & 1.35 & $-$1.66 \\
 B2782 &   NGC6752-15 & 19:10:49 & $-$60:01:55 & 13.73 & 4850 & 2.19 & 1.36 & $-$1.63 \\
 B4446 &   NGC6752-16 & 19:10:15 & $-$59:59:14 & 13.78 & 4906 & 2.24 & 1.33 & $-$1.63 \\
 B1113 &   NGC6752-19 & 19:11:23 & $-$59:59:40 & 13.96 & 4928 & 2.32 & 1.33 & $-$1.68 \\
\ldots &   NGC6752-20 & 19:10:36 & $-$59:56:08 & 13.98 & 4929 & 2.33 & 1.32 & $-$1.63 \\
\ldots &   NGC6752-21 & 19:11:13 & $-$60:02:30 & 14.02 & 4904 & 2.33 & 1.31 & $-$1.67 \\
 B1668 &   NGC6752-23 & 19:11:12 & $-$59:58:29 & 14.06 & 4916 & 2.35 & 1.25 & $-$1.66 \\
\ldots &   NGC6752-24 & 19:10:44 & $-$59:59:41 & 14.06 & 4948 & 2.37 & 1.16 & $-$1.71 \\
\ldots &   NGC6752-29 & 19:10:17 & $-$60:01:00 & 14.18 & 4950 & 2.42 & 1.31 & $-$1.69 \\
\ldots &   NGC6752-30 & 19:10:39 & $-$59:59:47 & 14.19 & 4943 & 2.42 & 1.26 & $-$1.64 \\
\hline
\end{tabular}
\end{minipage}
\end{table*}

Based on multi-band \textit{Hubble Space Telescope} (\textit{HST}) and
ground-based Str{\" o}mgren photometry, \citet{milone13} have identified three
populations on the main sequence, subgiant branch and red giant branch of NGC
6752. These populations, which we refer to as $a$, $b$ and $c$, exhibit
distinct chemical abundance patterns: population $a$ has a chemical composition
similar to that of field halo stars (e.g., high O and low Na); population $c$
is enhanced in N, Na and He ($\Delta Y \sim 0.03)$ and depleted C and O;
population $b$ has a chemical composition intermediate between populations $a$
and $c$ with slightly enhanced He ($\Delta Y \sim 0.01)$. Using the data from
\citet{milone13}, we can classify all program stars according to their
populations. In the relevant figures, stars of populations $a$, $b$ and $c$
are coloured green, magenta and blue, respectively. 

\subsection{Line List and Equivalent Width Measurements} 

\begin{figure}
\centering
      \includegraphics[width=.80\hsize]{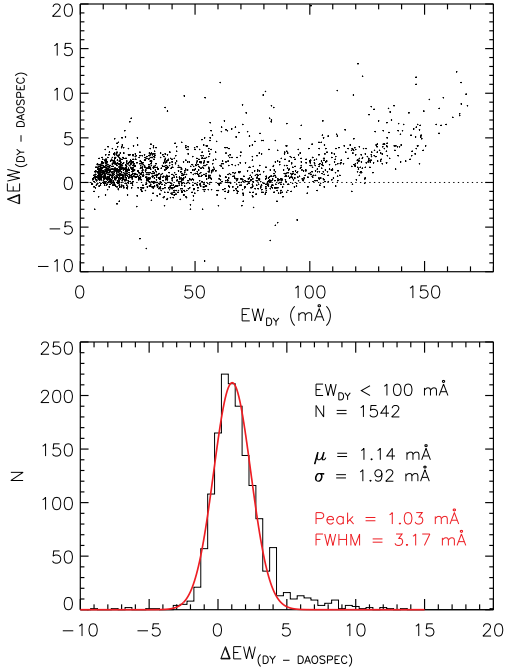} 
      \caption{Comparison of EWs measured using IRAF (DY) and DAOSPEC. The upper
panel shows all lines (N = 1,795). The lower panel shows the distribution of the
EW differences for the 1,542 lines with EW$_{\rm DY}$ $<$ 100 m\AA\ (i.e.,
measured using IRAF). We superimpose the Gaussian fit to the distribution and
write the relevant parameters associated with the fit as well as the mean and
dispersion.}
         \label{fig:ewcomp}
\end{figure}

The first step in our analysis was to measure equivalent widths (EWs) for a
large set of lines. The line list was taken primarily from \citet{gratton03}
and supplemented with laboratory measurements for \fei\ from the Oxford group
\citep{blackwell79feb,blackwell79fea,blackwell80fea,blackwell86fea,blackwell95fea},
laboratory measurements for \feii\ from \citet{biemont91} and for various
elements, the values taken from the references listed in \citet{yong05} (which
are also listed in Tables \ref{tab:ewtip} and \ref{tab:ewbump}).  We used the
DAOSPEC \citep{stetson08} software package to measure EWs in our program stars.
For the subset of lines we had previously measured using routines in
IRAF\footnote{IRAF (Image Reduction and Analysis Facility) is distributed by
the National Optical Astronomy Observatory, which is operated by the
Association of Universities for Research in Astronomy, Inc., under cooperative
agreement with the National Science Foundation.}, we compared those values with
the DAOSPEC measurements and found excellent agreement between the two sets of
EW measurements for lines having strengths less than $\sim$100m\AA\ (see Figure
\ref{fig:ewcomp}). For the 1,542 lines with EW $<$ 100~m\AA, we find a mean
difference EW(DY) $-$ EW(DAOSPEC) = 1.14 $\pm$ 0.05~m\AA\ ($\sigma$ =
1.92~m\AA). For our analysis, we adopted only lines with 5~m\AA\ $<$ EW $<$
100~m\AA\ as measured by DAOSPEC. A further requirement was that a given line
must be measured in every RGB tip star or every RGB bump star. That is, the
line list for the RGB tip sample was different from the line list for the RGB
bump sample, but for either sample of stars, each line was measured in every
star within a particular sample. Due to the lower quality spectra for the RGB
bump sample, we required lines to have EW $\ge$ 10~m\AA. The line list and EW
measurements for the RGB tip sample and for the RGB bump sample are presented
in Tables \ref{tab:ewtip} and \ref{tab:ewbump}, respectively. 

\begin{table*}
 \centering
 \begin{minipage}{140mm}
  \caption{Line List for the RGB Tip Stars\label{tab:ewtip}}
  \begin{tabular}{@{}cccrrrrrrc@{}}
  \hline
Wavelength & 
Species\footnote{The digits to the left of the decimal point are the atomic
number. The digit to the right of the decimal point is the ionization state
(``0'' = neutral, ``1'' = singly ionised).} & 
L.E.P & 
$\log gf$ & 
mg0\footnote{Star names are abbreviated. See Table \ref{tab:param} for the full
names.} &
mg1 & 
mg2 & 
mg3 & 
mg4 & 
Source\footnote{A = $\log gf$ values taken from \citet{yong05} where the references include 
\citet{denhartog03}, \citet{ivans01}, \citet{kurucz95}, 
\citet{prochaska00}, \citet{ramirez02}; 
B = \citet{gratton03};
C = Oxford group including 
\citet{blackwell79feb,blackwell79fea,blackwell80fea,blackwell86fea,blackwell95fea}; 
D = \citet{biemont91}} \\ 
\AA & 
 & 
eV & 
 & 
m\AA & 
m\AA & 
m\AA & 
m\AA & 
m\AA & 
 \\ 
(1) & 
(2) &
(3) &
(4) & 
(5) & 
(6) & 
(7) & 
(8) & 
(9) & 
(10) \\
\hline
6154.23 & 11.0 & 2.10 & $-$1.56 & 48.2 & 32.2 & 23.9 & 18.5 & 20.3 & A \\
6160.75 & 11.0 & 2.10 & $-$1.26 & 74.7 & 53.1 & 42.1 & 34.2 & 37.7 & A \\
5645.61 & 14.0 & 4.93 & $-$2.14 & 16.0 & 16.3 & 15.8 & 15.6 & 15.9 & A \\
5665.56 & 14.0 & 4.92 & $-$2.04 & 20.3 & 20.4 & 20.4 & 19.4 & 19.4 & B \\
5684.49 & 14.0 & 4.95 & $-$1.65 & 35.0 & 36.1 & 34.2 & 34.1 & 33.3 & B \\
\hline
\end{tabular}
\end{minipage}
\\ 
\vspace{5mm}
This table is published in its entirety in the electronic edition of the MNRAS.
A portion is shown here for guidance regarding its form and content. 
\end{table*}

\begin{table*}
 \centering
 \begin{minipage}{140mm}
  \caption{Line List for the RGB Bump Stars\label{tab:ewbump}.}
  \begin{tabular}{@{}cccrrrrrrc@{}}
  \hline
Wavelength & 
Species\footnote{The digits to the left of the decimal point are the atomic
number. The digit to the right of the decimal point is the ionization state
(``0'' = neutral, ``1'' = singly ionised).} & 
L.E.P & 
$\log gf$ & 
0\footnote{Star names are abbreviated. See Table \ref{tab:param} for the full
names.} &
1 & 
2 & 
3 & 
4 & 
Source\footnote{A = $\log gf$ values taken from \citet{yong05} where the references include 
\citet{denhartog03}, \citet{ivans01}, \citet{kurucz95}, 
\citet{prochaska00}, \citet{ramirez02}; 
B = \citet{gratton03};
C = Oxford group including 
\citet{blackwell79feb,blackwell79fea,blackwell80fea,blackwell86fea,blackwell95fea}; 
D = \citet{biemont91}} \\ 
\AA & 
 & 
eV & 
 & 
m\AA & 
m\AA & 
m\AA & 
m\AA & 
m\AA & 
 \\ 
(1) & 
(2) &
(3) &
(4) & 
(5) & 
(6) & 
(7) & 
(8) & 
(9) & 
(10) \\
\hline
5682.65 & 11.0 & 2.10 & $-$0.71 & 52.1 & 18.6 & 56.1 & 15.3 & 50.2 & A \\
5688.22 & 11.0 & 2.10 & $-$0.40 & 77.0 & 31.9 & 75.5 & 27.3 & 73.8 & A \\
5684.49 & 14.0 & 4.95 & $-$1.65 & 24.4 & 22.9 & 22.5 & 20.8 & 23.6 & B \\
5708.40 & 14.0 & 4.95 & $-$1.47 & 38.3 & 28.7 & 33.9 & 28.4 & 30.4 & B \\
5948.55 & 14.0 & 5.08 & $-$1.23 & 43.5 & 36.9 & 39.4 & 31.8 & 37.5 & A \\
\hline 
\end{tabular}
\end{minipage}
\\ 
\vspace{5mm}
This table is published in its entirety in the electronic edition of the MNRAS.
A portion is shown here for guidance regarding its form and content. 
\end{table*}

\subsection{Establishing Parameters for Reference Stars} 

In order to conduct the line-by-line strictly differential analysis, we needed
to adopt a reference star. The reference star parameters were determined in the
following manner. Note that since we did not know which reference stars would
be adopted, the procedure was applied to all stars. Following our previous
analyses of these spectra, effective temperatures, \teff, were derived from the
\citet{grundahl99} $uvby$ photometry using the \citet{alonso99}
\teff:colour:[Fe/H] relations. Surface gravities, \logg, were estimated using
\teff\ and the stellar luminosity. The latter value was determined by assuming
a mass of 0.84 M$_\odot$, a reddening $E(B-V)$ = 0.04 \citep{harris96} and
bolometric corrections taken from a 14 Gyr isochrone with [Fe/H] = $-$1.54 from
\citet{vandenberg00}. 

The model atmospheres used in the analysis were the one-dimensional,
plane-parallel, local thermodynamic equilibrium (LTE), $\alpha$-enhanced,
[$\alpha$/Fe] = +0.4, NEWODF grid of ATLAS9 models by \citet{castelli03}. We
used linear interpolation software (written by Dr Carlos Allende Prieto and
tested in \citealt{allende04}) to produce a particular model. (See
\citealt{meszaros13} for a discussion of interpolation of model atmospheres.)
Using the 2011 version of the stellar line analysis program MOOG
\citep{moog,sobeck11}, we computed the abundance for a given line.  The
microturbulent velocity, \vt, was set, in the usual way, by forcing the
abundances from \fei\ lines to have zero slope against the reduced equivalent
width, EW$_r$ = $\log (W_\lambda/\lambda)$. The metallicity was inferred from
\fei\ lines. We iterated this process until the inferred metallicity matched
the value adopted to generate the model atmosphere (this process usually
converged within three  iterations). (We exclude the RGB bump star NGC 6752-7
(B2438) due to its discrepant iron abundance, most likely resulting from a
photometric blend which affected the \teff\ and \logg\ values.) 
 
\subsection{Line-by-line Strictly Differential Stellar Parameters} 

Following \citet{melendez12}, we determined the stellar parameters using a
strictly differential line-by-line analysis between the program stars and a
reference star. Given the difference in \teff\ between the RGB tip and RGB bump
samples, we treated each sample separately. 

For the RGB tip stars, we selected NGC 6752-mg9 to be the reference star since
it had a \teff\ value close to the median for the RGB tip stars and the
O/Na/Mg/Al abundances were also close to the median values. These decisions
were motivated by the expectation that the errors in the derived stellar
parameters, and therefore errors in the chemical abundances, would increase if
there was a large difference in \teff\ between the program star and the
reference star.  Thus, we selected a star with \teff\ close to the median value
to minimise the difference in \teff\ between the program stars and the
reference star.  Similarly, we were concerned that large differences in the
abundances of O/Na/Mg/Al between the program star and the reference star could
increase the errors in the derived stellar parameters and chemical abundances.
Again, selecting the reference star to have O/Na/Mg/Al abundances close to the
median value minimises the abundance differences between the program stars and
the reference star. Application of a similar approach to the RGB bump sample
resulted in the selection of NGC 6752-11 as the reference star. 

To determine the stellar parameters for a program star, we generated a model
atmosphere with a particular combination of effective temperature (\teff),
surface gravity (\logg), microturbulent velocity (\vt) and metallicity, [m/H].
The initial guesses for these parameters came from the values in Section 2.3.
Using MOOG, we computed the abundances for \fei\ and \feii\ lines. We then
examined the {\it line-by-line Fe abundance differences}.  Adopting the
notation from \citet{melendez12}, the abundance difference (program star $-$
reference star) for a line is 
\begin{equation}
\delta A_i = A_i^{\rm program~star} - A_i^{\rm reference~star}. 
\end{equation} 

We examined the abundance differences for \fei\ as a function of lower
excitation potential. We forced excitation equilibrium by imposing the
following constraint 
\begin{equation}\label{eq:teff} 
\frac{\partial(\delta A_i^{\rm FeI})}{\partial(\chi_{\rm exc})} = 0. 
\end{equation}

Next, we considered the abundance differences for \fei\ as a function of
reduced equivalent width, EW$_r$, and imposed the following constraint
\begin{equation}\label{eq:vt}
\frac{\partial(\delta A_i^{\rm FeI})}{\partial({\rm EW}_r)} = 0. 
\end{equation}

For any species, \fei\ in this example, we then defined the average abundance
difference as 
\begin{equation}
\Delta^{\rm FeI} = \langle \delta A_i^{\rm FeI} \rangle = \frac{1}{N}  
\sum\limits_{i=1}^N \delta A_i^{\rm FeI} 
\end{equation}

Similarly, we defined the average \feii\ abundance as $\Delta^{\rm FeII}$ =
$\langle \delta A_i^{\rm FeII} \rangle$, and the relative ionization
equilibrium as 
\begin{equation}\label{eq:logg}
\Delta^{\rm FeI - FeII} = \Delta^{\rm FeI} - \Delta^{\rm FeII} = \langle \delta
A_i^{\rm FeI} \rangle - 
\langle \delta A_i^{\rm FeII} \rangle = 0. 
\end{equation} 
Unlike \citet{melendez12}, we did not take into account the relative
ionization equilibria for Cr and Ti, nor did we consider non-LTE effects for
any species. We note that while departures from LTE are expected for \fei\ for
metal-poor giants \citep{lind12}, the relative non-LTE effects across our range
of stellar parameters are vanishingly small. 

The final stellar parameters for a program star were obtained when equations
(\ref{eq:teff}), (\ref{eq:vt}) and (\ref{eq:logg}) were simultaneously
satisfied and the derived metallicity was identical to that used in generating
the model atmosphere. Regarding the latter criterion, we provide the following
example.  The metallicity of the reference star NGC 6752-mg9 was [Fe/H] =
$-$1.66 when adopting the \citet{asplund09} solar abundances and the
photometric stellar parameters described in Section 2.3 (see Table
\ref{tab:param}). For star NGC 6752-mg8, the average abundance difference for
\fei, and also \feii\ given equation (\ref{eq:logg}), was $\langle \delta
A_i^{\rm FeI} \rangle$ = +0.01 dex. Thus, the stellar parameters can only be
regarded as final if equations (\ref{eq:teff}), (\ref{eq:vt}) and
(\ref{eq:logg}) are satisfied and the model atmosphere is generated assuming a
global metallicity of [m/H] = [Fe/H]$_{\rm NGC 6752-mg9}$ + $\langle \delta
A_i^{\rm FeI} \rangle$ = $-$1.65. 

\begin{figure*}
\centering
      \includegraphics[width=.60\hsize]{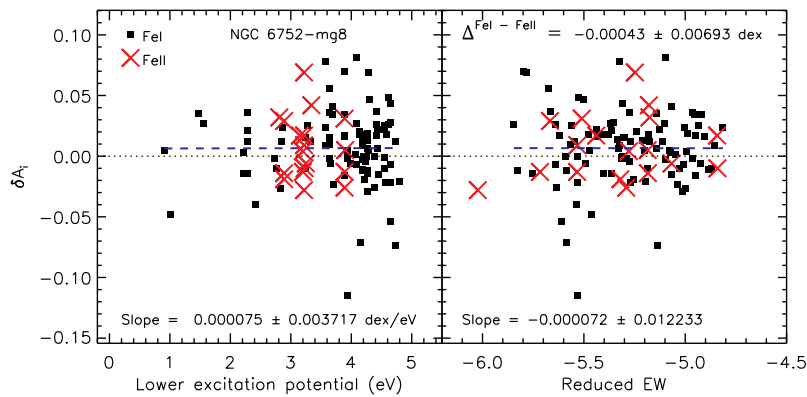} 
      \caption{Abundance differences, $\delta A_i$, for the RGB tip star NGC6752-mg8
(reference star NGC 6752-mg9) versus lower excitation potential (left) and
reduced EW (right). Values for \fei\ and \feii\ are shown as black squares and
red crosses, respectively. The blue dashed line in each panel is the linear
least squares fit to the data and we write the slope and associated uncertainty
in each panel. In the right panel, we also write $\Delta^{\rm FeI - FeII} =
\langle \delta A_i^{\rm FeI} \rangle - \langle \delta A_i^{\rm FeII} \rangle$.}
         \label{fig:paramtip}
\end{figure*}

\begin{figure*}
\centering
      \includegraphics[width=.60\hsize]{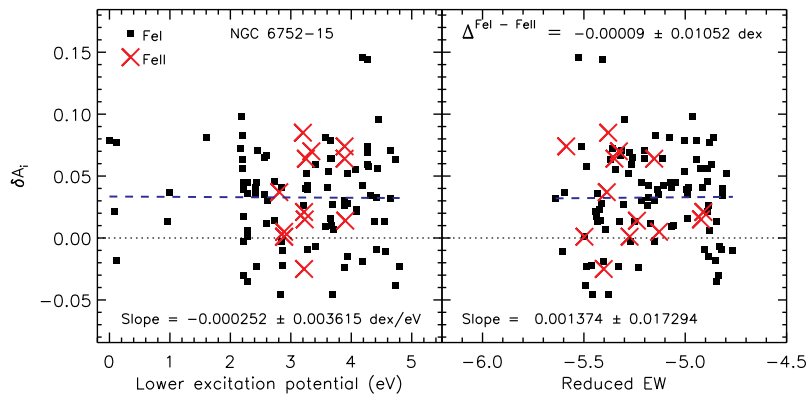} 
      \caption{Same as Figure \ref{fig:paramtip} but for the RGB bump star NGC
6752-15 (reference star NGC 6752-11).}
         \label{fig:parambum}
\end{figure*}

While equations (\ref{eq:teff}), (\ref{eq:vt}) and (\ref{eq:logg}) are
primarily sensitive to \teff, \vt\ and \logg, respectively, in practice, all
three equations are affected by small changes in any stellar parameter.
Derivation of these strictly differential stellar parameters required multiple
iterations (up to 20) where each iteration selected a single value for [m/H]
and five values for each parameter, \teff, \logg\ and \vt, in steps of 5~K,
0.05~dex and 0.05~\kms, respectively, i.e., 125 models per iteration. We then
examined the output from the 125 models to see whether equations (2), (3) and
(5) were simultaneously satisfied and whether the derived metallicity matched
that of the model atmosphere. If not, the best model was identified and we
repeated the process. If so, we conducted a final iteration in which we
selected a single value for [m/H] and tested 11 values for each parameter,
\teff, \logg\ and \vt, in steps of 1~K, 0.01~dex and 0.01~\kms, respectively,
i.e., 1,331 models in the final iteration using a smaller step size for each
parameter, and the best model was selected. As noted, this process was
performed separately for the RGB tip sample and for the RGB bump sample. The
strictly differential stellar parameters obtained using this pair of reference
stars (RGB tip = NGC 6752-mg9, RGB bump = NGC 6752-11) are presented in Table
\ref{tab:param1}.  (We exclude the RGB tip star NGC 6752-mg1 because the
stellar parameters did not converge.  Specifically, the best solution required
a value for \logg\ beyond the boundary of the \citet{castelli03} grid of model
atmospheres.) Figures \ref{fig:paramtip} and \ref{fig:parambum} provide
examples of $\delta A_i$, for \fei\ and \feii, versus lower excitation
potential and reduced EW for the strictly differential stellar parameters for a
representative RGB tip star and a representative RGB bump star, respectively.
That is, these figures show the results when equations (2), (3) and (5) are
simultaneously satisfied and the derived metallicity is the same as that used
to generate the model atmosphere. 

\begin{table*}
 \centering
 \begin{minipage}{140mm}
  \caption{Strictly Differential Stellar Parameters, and Uncertainties, When
Adopting the First Set of Reference Stars (RGB Tip = NGC 6752-mg9, RGB Bump =
NGC 6752-11). \label{tab:param1}} 
  \begin{tabular}{@{}lccccccc@{}}
  \hline
         Name & 
         \teff & 
         $\sigma$ & 
         \logg & 
         $\sigma$ & 
         \vt & 
         $\sigma$ & 
         [Fe/H] \\ 
          & 
         (K) & 
         (K) & 
         (cm s$^{-2}$) & 
         (cm s$^{-2}$) & 
         (\kms) & 
         (\kms) & 
          \\
         (1) & 
         (2) &
         (3) &
         (4) & 
         (5) & 
         (6) & 
         (7) & 
         (8) \\ 
\hline 
NGC6752-mg0  & 3919 & 20 & 0.16 & 0.01 & 2.24 & 0.05 & $-$1.69 \\
NGC6752-mg2  & 3938 & 22 & 0.23 & 0.01 & 2.13 & 0.05 & $-$1.67 \\
NGC6752-mg3  & 4066 & 19 & 0.53 & 0.01 & 1.93 & 0.04 & $-$1.65 \\
NGC6752-mg4  & 4081 & 18 & 0.54 & 0.01 & 1.90 & 0.04 & $-$1.65 \\
NGC6752-mg5  & 4100 & 17 & 0.56 & 0.01 & 1.93 & 0.04 & $-$1.66 \\
NGC6752-mg6  & 4151 & 19 & 0.65 & 0.01 & 1.88 & 0.04 & $-$1.63 \\
NGC6752-mg8  & 4284 & 14 & 0.93 & 0.01 & 1.73 & 0.04 & $-$1.65 \\
NGC6752-mg10 & 4291 & 12 & 0.92 & 0.01 & 1.70 & 0.03 & $-$1.66 \\
NGC6752-mg12 & 4315 & 13 & 0.96 & 0.01 & 1.76 & 0.04 & $-$1.66 \\
NGC6752-mg15 & 4339 & 13 & 1.01 & 0.01 & 1.76 & 0.04 & $-$1.66 \\
NGC6752-mg18 & 4380 & 15 & 1.07 & 0.01 & 1.71 & 0.04 & $-$1.66 \\
NGC6752-mg21 & 4437 & 13 & 1.16 & 0.01 & 1.69 & 0.05 & $-$1.65 \\
NGC6752-mg22 & 4444 & 14 & 1.19 & 0.01 & 1.71 & 0.04 & $-$1.64 \\
NGC6752-mg24 & 4505 & 17 & 1.30 & 0.01 & 1.72 & 0.07 & $-$1.68 \\
NGC6752-mg25 & 4471 & 15 & 1.24 & 0.01 & 1.74 & 0.07 & $-$1.69 \\
NGC6752-0    & 4706 & 12 & 1.85 & 0.01 & 1.44 & 0.02 & $-$1.65 \\
NGC6752-1    & 4719 & 11 & 1.94 & 0.01 & 1.37 & 0.02 & $-$1.65 \\
NGC6752-2    & 4739 & 12 & 1.95 & 0.01 & 1.35 & 0.02 & $-$1.66 \\
NGC6752-3    & 4749 & 13 & 2.00 & 0.01 & 1.34 & 0.02 & $-$1.73 \\
NGC6752-4    & 4794 & 13 & 2.08 & 0.01 & 1.37 & 0.02 & $-$1.66 \\
NGC6752-6    & 4795 & 11 & 2.10 & 0.01 & 1.32 & 0.02 & $-$1.64 \\
NGC6752-8    & 4930 & 15 & 2.29 & 0.01 & 1.31 & 0.03 & $-$1.67 \\
NGC6752-9    & 4795 & 21 & 2.09 & 0.01 & 1.40 & 0.04 & $-$1.73 \\
NGC6752-10   & 4811 & 10 & 2.11 & 0.01 & 1.35 & 0.02 & $-$1.67 \\
NGC6752-12   & 4822 & 13 & 2.15 & 0.01 & 1.34 & 0.02 & $-$1.68 \\
NGC6752-15   & 4830 & 12 & 2.23 & 0.01 & 1.34 & 0.02 & $-$1.65 \\
NGC6752-16   & 4875 & 15 & 2.24 & 0.01 & 1.31 & 0.03 & $-$1.66 \\
NGC6752-19   & 4892 & 12 & 2.32 & 0.01 & 1.31 & 0.02 & $-$1.71 \\
NGC6752-20   & 4899 & 12 & 2.32 & 0.01 & 1.30 & 0.02 & $-$1.65 \\
NGC6752-21   & 4884 & 14 & 2.32 & 0.01 & 1.30 & 0.03 & $-$1.69 \\
NGC6752-23   & 4912 & 12 & 2.33 & 0.01 & 1.25 & 0.02 & $-$1.67 \\
NGC6752-24   & 4911 & 17 & 2.39 & 0.01 & 1.14 & 0.03 & $-$1.74 \\
NGC6752-29   & 4923 & 13 & 2.40 & 0.01 & 1.30 & 0.02 & $-$1.71 \\
NGC6752-30   & 4919 & 12 & 2.47 & 0.01 & 1.24 & 0.02 & $-$1.66 \\
\hline
\end{tabular}
\end{minipage}
\end{table*}

\begin{figure}
\centering
      \includegraphics[width=.65\hsize]{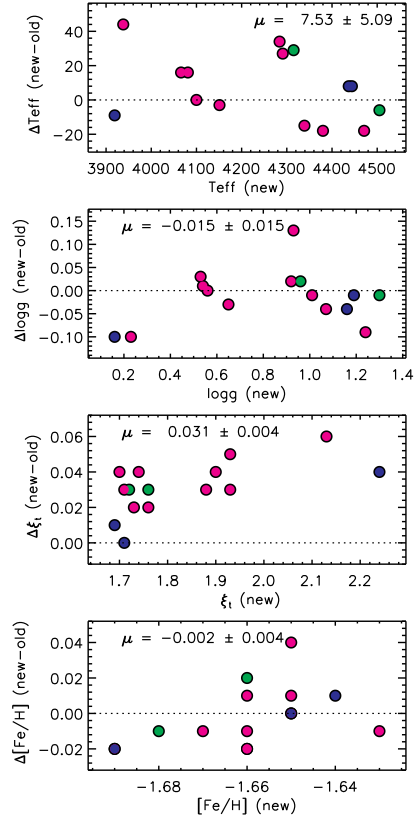} 
      \caption{Differences in \teff, \logg, \vt\ and [Fe/H] between the ``reference
star'' (old) and the ``strictly differential'' (new) stellar parameters for the
RGB tip sample (reference star is NGC 6752-mg9). The mean difference is written
in each panel. The green, magenta and blue colours represent populations $a$,
$b$ and $c$ from \citet{milone13} (see Section 2.1 for details).} 
         \label{fig:paramcomptip}
\end{figure}

\begin{figure}
\centering
      \includegraphics[width=.65\hsize]{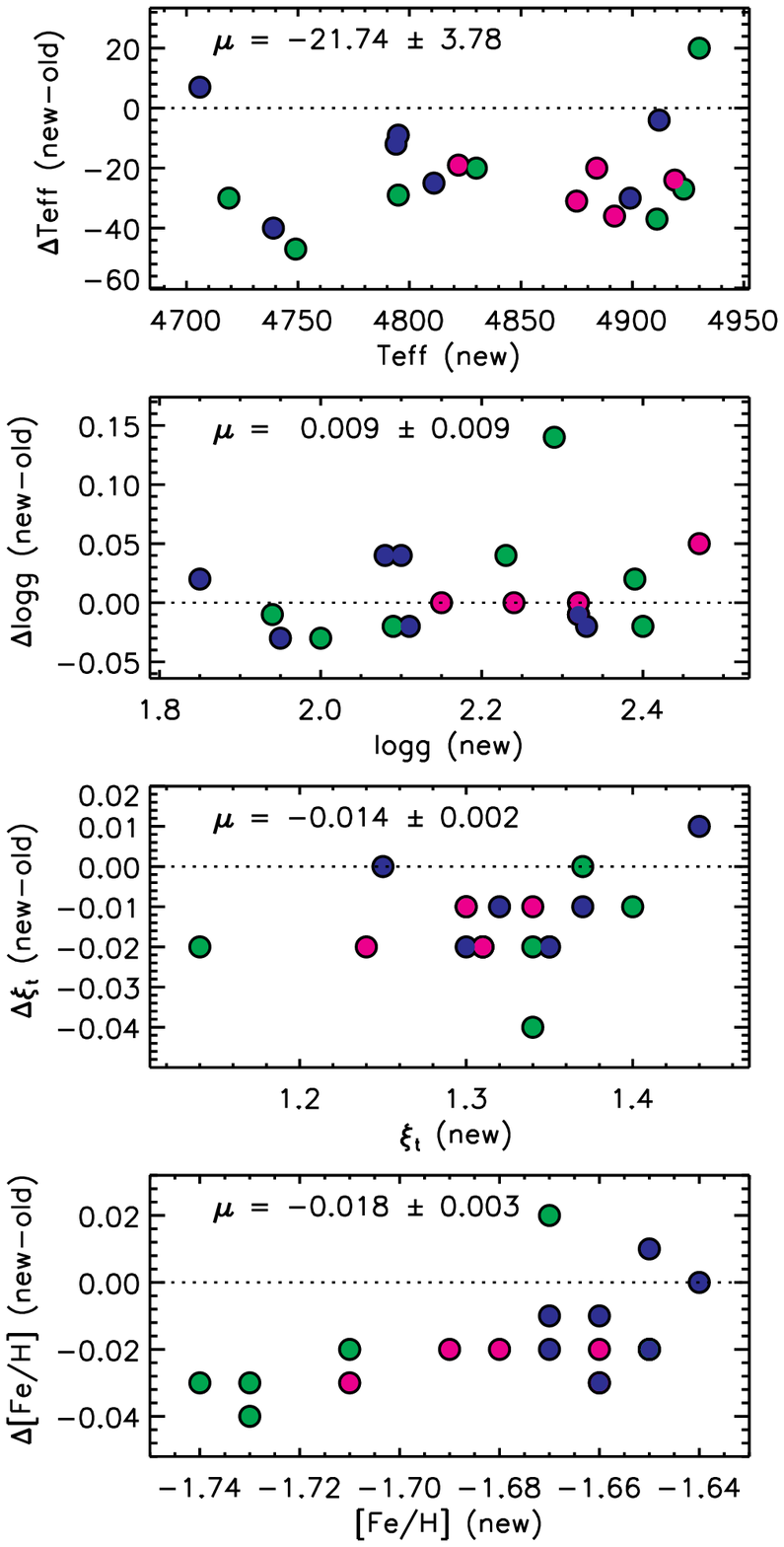} 
      \caption{Same as Figure \ref{fig:paramcomptip} but for the RGB bump
sample (reference star is NGC 6752-11).}
         \label{fig:paramcompbum}
\end{figure}

In Figures \ref{fig:paramcomptip} and \ref{fig:paramcompbum} we compare the
``reference star'' stellar parameters (described in Sec 2.3) and the ``strictly
differential'' stellar parameters (described above) for the RGB tip and RGB
bump samples, respectively, using the reference stars noted above. For the RGB
tip sample, the average difference between the ``reference star'' and
``strictly differential'' values for \teff, \logg, \vt\ and [Fe/H] are very
small; 7.53 K $\pm$ 5.09 K, $-$0.015 dex (cgs) $\pm$ 0.015 dex (cgs), 0.031
\kms\ $\pm$ 0.004 \kms\ and $-$0.002 dex $\pm$ 0.004 dex, respectively.
Comparably small differences in stellar parameters are obtained for the RGB
bump sample.  Therefore, an essential point we make here is that {\it the
strictly differential stellar parameters do not involve any substantial change
for any parameter, relative to the ``reference star'' stellar parameters}. For
\teff, the changes are within the uncertainties of the photometry. 

\subsection{Chemical Abundances} 

Having obtained the strictly differential stellar parameters, we computed the
abundances for the following species in every program star; Na, Si, Ca, \tii,
\tiii, \cri, \crii, Ni, Y, La, Nd and Eu. For the elements La and Eu, we used
spectrum synthesis and $\chi^2$ analysis of the 5380\AA\ and 6645\AA\ lines,
respectively, rather than an EW analysis since these lines are affected by
hyperfine splitting (HFS) and/or isotope shifts. We treated these lines
appropriately using the data from \citet{kurucz95} and for Eu, we adopted the
\citet{lodders03} solar isotope ratios. The $\log gf$ values for the La and Eu
lines were taken from \citet{lawler01la} and \citet{lawler01eu}, respectively. 

We used equation (1) to obtain the abundance difference (between the program
star and the reference star) for any line. For a particular species, X, the
average abundance difference is $\langle \delta A_i^{\rm X} \rangle$ which we
write as $\Delta^{\rm X}$, i.e., as defined in equation (4) above. In Tables 
\ref{tab:abuna} and \ref{tab:abunb}, we present the abundance differences for
each element in all program stars. In order to put these abundance differences
onto an absolute scale, in these tables we also provide the A(X) abundances for
the reference stars when using the stellar parameters in Table \ref{tab:param}.
The new [X/Fe] values are in very good agreement with our previously published
values \citep{grundahl02,yong03,yong05}, although we have not attempted to
reconcile the two sets of abundances. 

\begin{table*}
 \centering
 \begin{minipage}{170mm}
  \caption{Differential Abundances (Fe, Na, Si, Ca and Ti) When Adopting the
First Set of Reference Stars (RGB Tip = NGC 6752-mg9, RGB Bump = NGC
6752-11).\label{tab:abuna}} 
  \begin{tabular}{@{}lrcrcrcrcrcrc@{}}
  \hline
Star & 
$\Delta^{\rm Fe}$ &
$\sigma$ & 
$\Delta^{\rm Na}$ &
$\sigma$ & 
$\Delta^{\rm Si}$ &
$\sigma$ & 
$\Delta^{\rm Ca}$ &
$\sigma$ & 
$\Delta^{\rm TiI}$ &
$\sigma$ & 
$\Delta^{\rm TiII}$ &
$\sigma$ \\
(1) & 
(2) &
(3) &
(4) & 
(5) & 
(6) & 
(7) & 
(8) & 
(9) &
(10) & 
(11) & 
(12) & 
(13) \\
\hline 
NGC6752-mg0  & $-$0.029 & 0.010 &    0.387 & 0.016 &    0.038 & 0.015 & $-$0.023 & 0.033 &    0.021 & 0.020 & $-$0.024 & 0.035 \\
NGC6752-mg2  & $-$0.011 & 0.011 & $-$0.014 & 0.008 &    0.039 & 0.010 & $-$0.021 & 0.049 &    0.050 & 0.024 &    0.036 & 0.039 \\
NGC6752-mg3  &    0.007 & 0.015 & $-$0.027 & 0.005 &    0.007 & 0.009 & $-$0.003 & 0.045 &    0.020 & 0.017 &    0.043 & 0.041 \\
NGC6752-mg4  &    0.010 & 0.014 &    0.041 & 0.010 &    0.030 & 0.010 &    0.008 & 0.038 &    0.023 & 0.012 &    0.043 & 0.045 \\
NGC6752-mg5  &    0.005 & 0.008 &    0.052 & 0.008 &    0.015 & 0.008 &    0.001 & 0.015 &    0.006 & 0.012 &    0.038 & 0.035 \\
NGC6752-mg6  &    0.032 & 0.009 & $-$0.123 & 0.002 &    0.042 & 0.009 &    0.049 & 0.040 &    0.052 & 0.011 &    0.095 & 0.065 \\
NGC6752-mg8  &    0.007 & 0.010 &    0.036 & 0.015 & $-$0.001 & 0.017 &    0.004 & 0.024 &    0.008 & 0.023 &    0.029 & 0.019 \\
NGC6752-mg10 &    0.007 & 0.010 &    0.013 & 0.004 & $-$0.004 & 0.007 & $-$0.005 & 0.017 & $-$0.023 & 0.010 &    0.027 & 0.033 \\
NGC6752-mg12 &    0.002 & 0.010 & $-$0.342 & 0.004 & $-$0.021 & 0.007 & $-$0.017 & 0.016 &    0.006 & 0.007 & $-$0.007 & 0.030 \\
NGC6752-mg15 & $-$0.001 & 0.009 &    0.044 & 0.009 & $-$0.008 & 0.010 & $-$0.008 & 0.011 & $-$0.009 & 0.007 &    0.009 & 0.023 \\
NGC6752-mg18 & $-$0.002 & 0.010 & $-$0.094 & 0.004 & $-$0.006 & 0.009 & $-$0.016 & 0.017 & $-$0.018 & 0.007 &    0.044 & 0.033 \\
NGC6752-mg21 &    0.018 & 0.009 &    0.282 & 0.009 &    0.043 & 0.009 &    0.032 & 0.013 & $-$0.012 & 0.009 &    0.057 & 0.031 \\
NGC6752-mg22 &    0.014 & 0.009 &    0.323 & 0.008 &    0.030 & 0.010 &    0.017 & 0.011 & $-$0.012 & 0.010 &    0.012 & 0.031 \\
NGC6752-mg24 & $-$0.023 & 0.016 & $-$0.345 & 0.035 & $-$0.049 & 0.009 & $-$0.040 & 0.012 & $-$0.034 & 0.009 &    0.047 & 0.059 \\
NGC6752-mg25 & $-$0.027 & 0.010 & $-$0.139 & 0.025 & $-$0.008 & 0.010 & $-$0.026 & 0.023 & $-$0.045 & 0.009 & $-$0.023 & 0.039 \\
NGC6752-0    &    0.030 & 0.010 &    0.335 & 0.033 &    0.096 & 0.019 &    0.050 & 0.010 &    0.023 & 0.011 &    0.052 & 0.012 \\
NGC6752-1    &    0.025 & 0.009 & $-$0.366 & 0.020 & $-$0.008 & 0.013 &    0.031 & 0.010 &    0.003 & 0.011 &    0.034 & 0.012 \\
NGC6752-2    &    0.020 & 0.008 &    0.384 & 0.015 &    0.055 & 0.012 &    0.038 & 0.008 & $-$0.001 & 0.008 &    0.031 & 0.014 \\
NGC6752-3    & $-$0.049 & 0.012 & $-$0.444 & 0.016 & $-$0.044 & 0.007 & $-$0.044 & 0.009 & $-$0.052 & 0.013 & $-$0.036 & 0.017 \\
NGC6752-4    &    0.017 & 0.015 &    0.352 & 0.021 &    0.026 & 0.021 &    0.065 & 0.011 &    0.007 & 0.013 &    0.034 & 0.017 \\
NGC6752-6    &    0.036 & 0.014 &    0.262 & 0.017 &    0.032 & 0.008 &    0.060 & 0.011 &    0.027 & 0.013 &    0.042 & 0.014 \\
NGC6752-8    &    0.010 & 0.014 & $-$0.323 & 0.012 & $-$0.045 & 0.017 &    0.027 & 0.010 &    0.030 & 0.012 &    0.018 & 0.013 \\
NGC6752-9    & $-$0.048 & 0.025 & $-$0.396 & 0.056 & $-$0.049 & 0.011 & $-$0.038 & 0.013 & $-$0.062 & 0.016 & $-$0.045 & 0.018 \\
NGC6752-10   &    0.013 & 0.011 &    0.357 & 0.020 &    0.016 & 0.012 &    0.039 & 0.014 &    0.007 & 0.019 &    0.032 & 0.014 \\
NGC6752-12   &    0.000 & 0.013 & $-$0.065 & 0.009 & $-$0.012 & 0.016 &    0.003 & 0.010 & $-$0.023 & 0.013 &    0.027 & 0.016 \\
NGC6752-15   &    0.033 & 0.012 & $-$0.355 & 0.075 & $-$0.002 & 0.012 &    0.022 & 0.011 & $-$0.006 & 0.015 &    0.042 & 0.015 \\
NGC6752-16   &    0.021 & 0.016 &    0.091 & 0.014 & $-$0.005 & 0.018 &    0.008 & 0.011 &    0.001 & 0.015 &    0.007 & 0.016 \\
NGC6752-19   & $-$0.029 & 0.012 & $-$0.190 & 0.008 & $-$0.048 & 0.010 & $-$0.029 & 0.008 & $-$0.046 & 0.011 & $-$0.024 & 0.012 \\
NGC6752-20   &    0.029 & 0.012 &    0.454 & 0.015 &    0.031 & 0.015 &    0.051 & 0.009 &    0.020 & 0.013 &    0.037 & 0.012 \\
NGC6752-21   & $-$0.007 & 0.013 & $-$0.063 & 0.003 & $-$0.019 & 0.018 &    0.010 & 0.011 & $-$0.010 & 0.014 &    0.011 & 0.013 \\
NGC6752-23   &    0.016 & 0.012 &    0.272 & 0.019 &    0.032 & 0.012 &    0.033 & 0.009 & $-$0.002 & 0.013 &    0.024 & 0.015 \\
NGC6752-24   & $-$0.058 & 0.016 & $-$0.408 & 0.010 & $-$0.107 & 0.020 & $-$0.048 & 0.015 & $-$0.078 & 0.011 & $-$0.081 & 0.018 \\
NGC6752-29   & $-$0.026 & 0.012 & $-$0.421 & 0.032 & $-$0.101 & 0.020 & $-$0.025 & 0.009 & $-$0.064 & 0.021 & $-$0.043 & 0.012 \\
NGC6752-30   &    0.025 & 0.011 & $-$0.161 & 0.010 & $-$0.007 & 0.013 &    0.056 & 0.012 &    0.003 & 0.015 &    0.051 & 0.014 \\
\hline
\end{tabular}
\end{minipage}
\\
In order to place the above values onto an absolute scale, the absolute
abundances we obtain for the reference stars are given below. We caution,
however, that the absolute scale has not been critically evaluated (see Section
2.5 for more details). 
\\ 
NGC6752-mg9: 
A(Fe) = 5.85, 
A(Na) = 4.86, 
A(Si) = 6.23, 
A(Ca) = 4.99, 
A(\tii) = 3.54, 
A(\tiii) = 3.59. 
\\ 
NGC6752-11: 
A(Fe) = 5.84, 
A(Na) = 4.84, 
A(Si) = 6.24, 
A(Ca) = 4.97, 
A(\tii) = 3.50, 
A(\tiii) = 3.72. 
\end{table*}

\begin{table*}
 \centering
 \begin{minipage}{190mm}
  \caption{Differential Abundances (Cr, Ni, Y, La, Nd and Eu) When Adopting the
First Set of Reference Stars (RGB Tip = NGC 6752-mg9, RGB Bump = NGC
6752-11).\label{tab:abunb}} 
  \begin{tabular}{@{}lrcrcrcrcrcrcrcrcrcrcrcrcrc@{}}
  \hline
Star & 
$\Delta^{\rm CrI}$ &
$\sigma$ & 
$\Delta^{\rm CrII}$ &
$\sigma$ & 
$\Delta^{\rm Ni}$ &
$\sigma$ & 
$\Delta^{\rm Y}$ &
$\sigma$ & 
$\Delta^{\rm La}$ &
$\sigma$ & 
$\Delta^{\rm Nd}$ &
$\sigma$ & 
$\Delta^{\rm Eu}$ &
$\sigma$ \\ 
(1) & 
(2) &
(3) &
(4) & 
(5) & 
(6) & 
(7) & 
(8) & 
(9) &
(10) & 
(11) & 
(12) & 
(13) & 
(14) & 
(15) \\
\hline 
NGC6752-mg0  &    0.013 & 0.059 &    0.018 & 0.077 & $-$0.030 & 0.023 &    0.022 & 0.037 &    0.028 & 0.013 & $-$0.011 & 0.042 & $-$0.002 &  0.012 \\
NGC6752-mg2  &    0.053 & 0.087 &    0.068 & 0.074 & $-$0.000 & 0.021 &    0.087 & 0.045 &    0.081 & 0.017 &    0.051 & 0.058 & $-$0.012 &  0.013 \\
NGC6752-mg3  &    0.042 & 0.046 &    0.042 & 0.035 &    0.005 & 0.023 &    0.074 & 0.036 &    0.106 & 0.016 &    0.046 & 0.061 &    0.063 &  0.013 \\
NGC6752-mg4  &    0.050 & 0.046 &    0.055 & 0.033 &    0.013 & 0.019 &    0.075 & 0.024 &    0.073 & 0.015 &    0.058 & 0.042 &    0.056 &  0.014 \\
NGC6752-mg5  &    0.034 & 0.037 &    0.023 & 0.029 & $-$0.001 & 0.011 &    0.006 & 0.035 &    0.140 & 0.016 &    0.029 & 0.026 &    0.027 &  0.014 \\
NGC6752-mg6  &    0.028 & 0.042 &    0.044 & 0.024 &    0.038 & 0.023 &    0.098 & 0.028 &    0.109 & 0.017 &    0.067 & 0.053 &    0.060 &  0.014 \\
NGC6752-mg8  & $-$0.029 & 0.035 & $-$0.095 & 0.085 &    0.007 & 0.014 &    0.015 & 0.013 &    0.087 & 0.016 &    0.026 & 0.016 &    0.053 &  0.016 \\
NGC6752-mg10 & $-$0.009 & 0.022 & $-$0.055 & 0.074 & $-$0.001 & 0.012 &    0.079 & 0.020 &    0.020 & 0.017 &    0.019 & 0.025 & $-$0.032 &  0.016 \\
NGC6752-mg12 & $-$0.005 & 0.013 & $-$0.014 & 0.006 &    0.003 & 0.008 & $-$0.006 & 0.020 & $-$0.036 & 0.016 &    0.000 & 0.021 &    0.013 &  0.016 \\
NGC6752-mg15 & $-$0.027 & 0.011 & $-$0.019 & 0.014 & $-$0.006 & 0.007 & $-$0.001 & 0.004 &    0.042 & 0.016 &    0.015 & 0.013 & $-$0.013 &  0.014 \\
NGC6752-mg18 & $-$0.026 & 0.016 & $-$0.032 & 0.014 & $-$0.007 & 0.010 &    0.014 & 0.026 &    0.005 & 0.018 & $-$0.011 & 0.028 &    0.007 &  0.017 \\
NGC6752-mg21 & $-$0.003 & 0.023 & $-$0.021 & 0.012 & $-$0.002 & 0.008 &    0.068 & 0.023 &    0.059 & 0.017 &    0.010 & 0.022 & $-$0.037 &  0.017 \\
NGC6752-mg22 & $-$0.017 & 0.042 &    0.007 & 0.039 &    0.009 & 0.009 &    0.047 & 0.018 &    0.049 & 0.017 &    0.013 & 0.016 &    0.008 &  0.018 \\
NGC6752-mg24 & $-$0.033 & 0.013 & $-$0.060 & 0.013 & $-$0.024 & 0.008 & $-$0.062 & 0.015 & $-$0.005 & 0.016 & $-$0.032 & 0.018 &    0.018 &  0.018 \\
NGC6752-mg25 & $-$0.023 & 0.021 & $-$0.046 & 0.014 & $-$0.043 & 0.010 & $-$0.038 & 0.018 &    0.108 & 0.015 & $-$0.051 & 0.026 &    0.003 &  0.018 \\
NGC6752-0    &    0.058 & 0.012 &    0.112 & 0.053 &    0.020 & 0.009 &    0.044 & 0.018 &    0.018 & 0.012 &    0.018 & 0.015 &    0.123 &  0.024 \\
NGC6752-1    &    0.037 & 0.014 &    0.077 & 0.060 &    0.010 & 0.014 &    0.026 & 0.027 & $-$0.060 & 0.012 & $-$0.009 & 0.025 & $-$0.068 &  0.026 \\
NGC6752-2    &    0.009 & 0.012 &    0.038 & 0.005 & $-$0.003 & 0.008 & $-$0.017 & 0.023 &    0.032 & 0.011 & $-$0.009 & 0.029 &    0.180 &  0.023 \\
NGC6752-3    & $-$0.053 & 0.023 & $-$0.053 & 0.029 & $-$0.057 & 0.013 & $-$0.143 & 0.009 & $-$0.039 & 0.012 & $-$0.110 & 0.025 &    0.089 &  0.025 \\
NGC6752-4    &    0.014 & 0.023 &    0.062 & 0.046 &    0.003 & 0.012 &    0.018 & 0.022 &    0.009 & 0.010 & $-$0.014 & 0.027 &    0.328 &  0.025 \\
NGC6752-6    &    0.038 & 0.027 &    0.068 & 0.052 &    0.004 & 0.012 &    0.005 & 0.025 &    0.027 & 0.013 &    0.041 & 0.035 &    0.208 &  0.025 \\
NGC6752-8    &    0.019 & 0.016 &    0.061 & 0.055 & $-$0.004 & 0.008 & $-$0.026 & 0.026 &    0.064 & 0.010 &    0.033 & 0.014 &    0.179 &  0.029 \\
NGC6752-9    & $-$0.039 & 0.026 &    0.028 & 0.044 & $-$0.054 & 0.016 & $-$0.089 & 0.012 & $-$0.014 & 0.011 & $-$0.064 & 0.023 &    0.149 &  0.025 \\
NGC6752-10   &    0.029 & 0.022 &    0.016 & 0.022 & $-$0.016 & 0.014 &    0.016 & 0.013 &    0.076 & 0.012 & $-$0.013 & 0.025 &    0.185 &  0.029 \\
NGC6752-12   &    0.004 & 0.021 &    0.075 & 0.065 & $-$0.016 & 0.010 & $-$0.097 & 0.021 & $-$0.006 & 0.011 & $-$0.020 & 0.032 &    0.008 &  0.028 \\
NGC6752-15   &    0.024 & 0.021 &    0.070 & 0.021 &    0.005 & 0.013 & $-$0.046 & 0.026 & $-$0.005 & 0.011 & $-$0.010 & 0.025 & $-$0.082 &  0.034 \\
NGC6752-16   &    0.016 & 0.019 &    0.012 & 0.024 &    0.007 & 0.013 & $-$0.048 & 0.015 &    0.031 & 0.013 &    0.045 & 0.031 & $-$0.001 &  0.039 \\
NGC6752-19   & $-$0.036 & 0.021 &    0.016 & 0.048 & $-$0.052 & 0.010 & $-$0.107 & 0.013 &    0.018 & 0.011 & $-$0.049 & 0.024 &    0.004 &  0.042 \\
NGC6752-20   &    0.024 & 0.018 &    0.038 & 0.019 &    0.007 & 0.007 &    0.012 & 0.014 &    0.054 & 0.012 &    0.011 & 0.026 &    0.057 &  0.042 \\
NGC6752-21   & $-$0.014 & 0.018 &    0.052 & 0.025 & $-$0.032 & 0.009 & $-$0.013 & 0.015 &    0.087 & 0.011 & $-$0.023 & 0.019 & $-$0.032 &  0.039 \\
NGC6752-23   &    0.006 & 0.025 &    0.102 & 0.036 & $-$0.026 & 0.010 &    0.016 & 0.010 & $-$0.028 & 0.011 & $-$0.004 & 0.011 & $-$0.033 &  0.040 \\
NGC6752-24   & $-$0.056 & 0.019 & $-$0.031 & 0.020 & $-$0.089 & 0.010 & $-$0.135 & 0.018 & $-$0.050 & 0.012 & $-$0.075 & 0.016 &    0.141 &  0.050 \\
NGC6752-29   & $-$0.036 & 0.020 &    0.051 & 0.042 & $-$0.056 & 0.011 & $-$0.082 & 0.022 & $-$0.094 & 0.012 & $-$0.054 & 0.021 &    0.062 &  0.033 \\
NGC6752-30   &    0.029 & 0.016 &    0.048 & 0.037 & $-$0.007 & 0.010 &    0.000 & 0.032 &    0.047 & 0.011 &    0.025 & 0.017 &    0.235 &  0.031 \\
\hline
\end{tabular}
\end{minipage}
\\
In order to place the above values onto an absolute scale, the absolute
abundances we obtain for the reference stars are given below. We caution,
however, that the absolute scale has not been critically evaluated (see Section
2.5 for more details). 
\\ 
NGC6752-mg9: 
A(\cri) = 3.99, 
A(\crii) = 4.10, 
A(Ni) = 4.56, 
A(Y) = 0.67, 
A(La) = $-$0.39, 
A(Nd) = 0.06, 
A(Eu) = $-$0.75. 
\\ 
NGC6752-11: 
A(\cri) = 3.84, 
A(\crii) = 4.12, 
A(Ni) = 4.54, 
A(Y) = 0.66, 
A(La) = $-$0.29, 
A(Nd) = 0.06, 
A(Eu) = $-$0.80. 
\end{table*}

For Na, the range in abundance is 0.90 dex, in good agreement with our
previously published values. We did not attempt to re-measure the abundances of
other the light elements, O, Mg and Al, as multiple lines could not be
measured in all stars. Additionally, given the well established correlations
between the abundances of these elements, we believe that Na provides a
reliable picture of the light element abundance variations in this cluster. The
interested reader can find our abundances for N, O, Mg and Al in
\citet{grundahl02} and \citet{yong03,yong08nh}. (C measurements in the RGB bump
sample are ongoing and will be presented in a future work.) 

\begin{figure}
\centering
      \includegraphics[width=1.0\hsize]{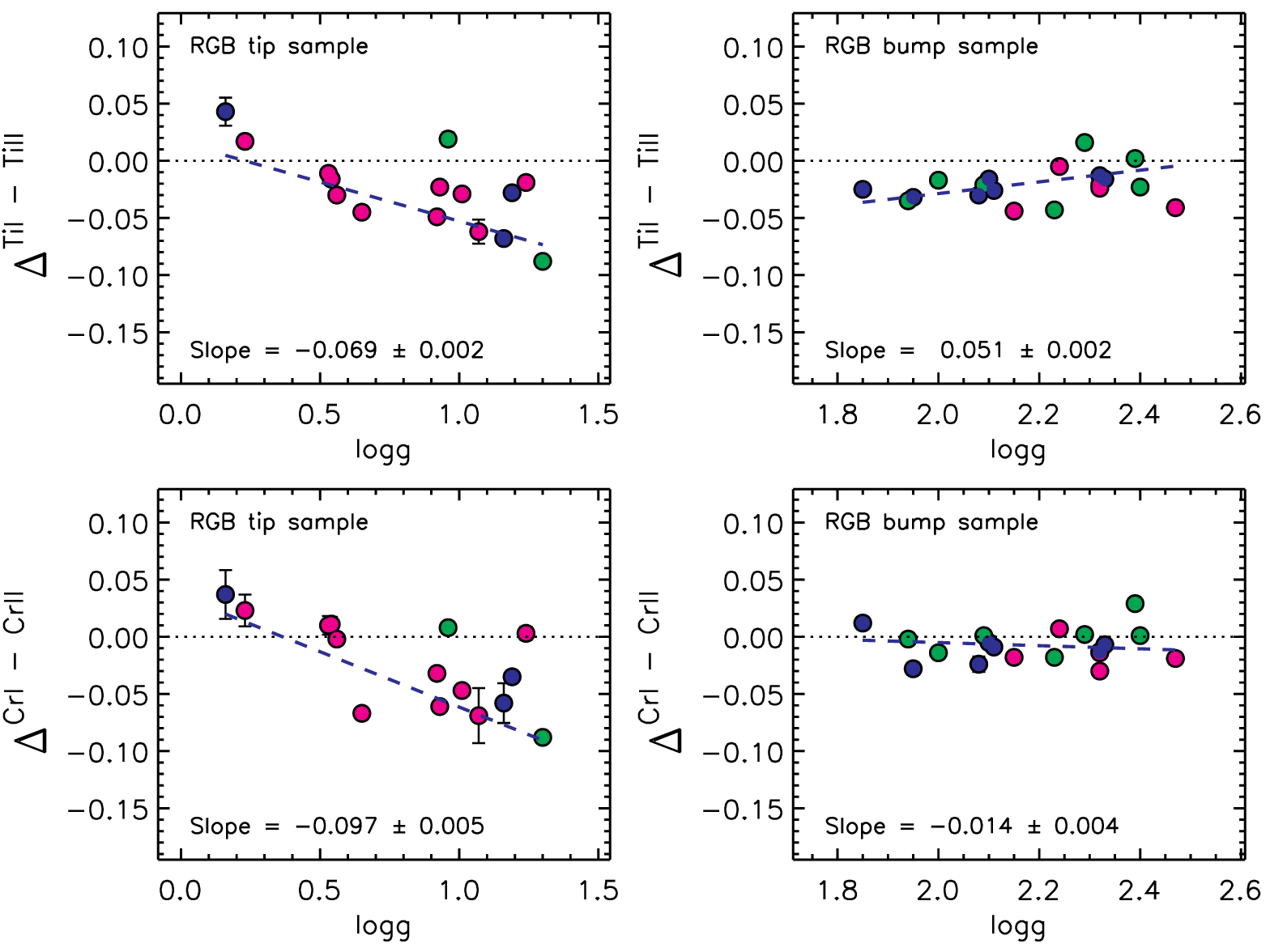} 
      \caption{$\Delta^{\rm TiI - TiII}$ (upper panels) and $\Delta^{\rm CrI - CrII}$
(lower panels) for the RGB tip star sample (left panels) and the RGB bump star
sample (right panels). (These results are obtained when using the reference
stars RGB tip = NGC 6752-mg9 and RGB bump = NGC 6752-11.) The colours are the
same as in Figure \ref{fig:paramcomptip}.}
         \label{fig:ticr}
\end{figure}

As mentioned, \citet{melendez12} considered the relative ionization equilibria
for Ti and Cr when establishing the strictly differential stellar parameters.
Having measured the Ti and Cr abundances from neutral and ionised lines, we are
now in a position to examine 
$\Delta^{\rm TiI - TiII} = \langle \delta A_i^{\rm TiI} \rangle - 
\langle \delta A_i^{\rm TiII} \rangle$ and 
$\Delta^{\rm CrI - CrII} = \langle \delta A_i^{\rm CrI} \rangle - 
\langle \delta A_i^{\rm CrII} \rangle$. In Figure \ref{fig:ticr}, we plot
$\Delta^{\rm TiI - TiII}$ and $\Delta^{\rm CrI - CrII}$ versus \logg\ for both
samples of stars. In this figure, it is clear that ionization equilibrium is
not obtained for Ti or Cr and that there are trends between $\Delta^{\rm TiI -
TiII}$ vs.\ \logg\ and $\Delta^{\rm CrI - CrII}$ vs.\ \logg. Nevertheless, we
are satisfied with our approach which used only Fe lines to establish the
differential stellar parameters. We expect that inclusion of Ti and Cr
ionization equilibrium would have resulted in very small adjustments to the
stellar parameters and to the differential chemical abundances. Finally, as it
will be shown later, Ti and Cr have considerably higher uncertainties such that
it may be better to rely only upon Fe for ionization balance. 

\subsection{Error Analysis}

To determine the errors in the stellar parameters, we adopted the following
approach. For \teff, we determined the formal uncertainty in the slope between
$\delta A_i^{\rm FeI}$ and the lower excitation potential. We then adjusted
\teff\ until the formal slope matched the error. The difference between the new
\teff\ and the original value is $\sigma$\teff. For the RGB tip and RGB bump
stars, the average values of $\sigma$\teff\ were 7.53 K and 21.74 K,
respectively. For \logg, we added the standard error of the mean for
$\Delta^{\rm FeI}$ and $\Delta^{\rm FeII}$ in quadrature and then adjusted
\logg\ until the quantity $\Delta^{\rm FeI - FeII}$, from equation
(\ref{eq:logg}) above, was equal to this value. The difference between the new
\logg\ and the original value is $\sigma$\logg. For the RGB tip and RGB bump
stars, the average values of $\sigma$\logg\ were 0.015 dex and 0.009 dex,
respectively. For \vt, we measured
the formal uncertainty in the slope between $\delta A_i^{\rm FeI}$ and the
reduced equivalent width.  We adjusted \vt\ until the formal slope was equal to
this value. The difference between the new and old values is $\sigma$\vt.
Average values for $\sigma$\vt\ for the RGB tip and RGB bump samples were 0.031
\kms\ and 0.018 \kms, respectively. 

Uncertainties in the element abundance measurements were obtained following the
formalism given in \citet{johnson02}, which we repeat here for convenience, and
we note that this approach is very similar to that of \citet{mcwilliam95} and
\citet{barklem05}. 
\begin{eqnarray}
\sigma^2_{\rm{log}\epsilon}= \sigma^2_{\rm rand} +
\left({\partial \rm{log}\epsilon\over\partial T}\right)^2 
\sigma^2_T + 
\left({\partial \rm{log}\epsilon\over\partial \mlogg}\right)^2 
\sigma^2_{\mlogg}  +
\nonumber\\ 
\left({\partial \rm{log}\epsilon\over\partial \xi}\right)^2 \sigma^2_{\xi} + 
2\biggl[\left({\partial \rm{log}\epsilon\over\partial T}\right)
\left({\partial \rm{log}\epsilon
\over\partial \mlogg}\right)\sigma_{T\mlogg} + 
\nonumber\\
 \left({\partial 
\rm{log}\epsilon\over\partial \xi}\right) \left({\partial \rm{log}\epsilon\over\partial 
\mlogg}\right) \sigma_{\mlogg \xi} +
 \left({\partial \rm{log}\epsilon\over\partial \xi}\right)\left({\partial 
\rm{log}\epsilon\over T}\right) \sigma_{\xi T} \biggr] 
\end{eqnarray}
The covariance terms, $\sigma_{T\mlogg}$, $\sigma_{\mlogg \xi}$ and
$\sigma_{\xi T}$, were computed using the approach of \citet{johnson02}.  These
abundance uncertainties are included in Tables \ref{tab:abuna},
\ref{tab:abunb}, \ref{tab:abun2a} and \ref{tab:abun2b}.  For La and Eu, the
abundances were obtained from a single line. For these lines, we adopt the
1$\sigma$ fitting error from the $\chi^2$ analysis in place of the random
error term, $\sigma_{\rm rand}$ (standard error of the mean).  We note that
these formal uncertainties, which take into account all covariance error terms,
are below 0.02 dex for many elements in many stars, reaching values as low as
$\sim$0.01 dex for a number of elements including Si, \tii, Ni and Fe. 

Note that in Figure \ref{fig:ticr}, we regard $\Delta^{\rm TiI - TiII}$ as an
abundance ratio between \tii\ and \tiii, and thus, we compute the error terms
according to the relevant equations in \citet{johnson02} which we again repeat
here for convenience. 
\begin{equation}
\sigma^2(A/B)=\sigma^2(A)+\sigma^2(B)-2 \sigma_{A,B}
\end{equation}
The covariance between two abundances is given by 
{\setlength
\arraycolsep{2pt}
\begin{eqnarray}
\sigma_{A,B} = \left({\Pd \ep _A\over\Pd T}\right)\left({\Pd \ep _{B}\over\Pd 
T}\right)\sigma^2_T  + \hspace{0.8in}  
\nonumber\\
\left({\Pd \ep _A\over\Pd \mlogg}\right)
\left({\Pd \ep _B\over\Pd \mlogg}\right) \sigma^2_{\mlogg}  +  
\left({\Pd \ep _A\over\Pd \xi}\right) \left({\Pd \ep _B\over\Pd \xi}\right)
\sigma^2_{\xi} 
\nonumber\\
+ \biggl[\left({\Pd \ep _{A}\over\Pd T}\right)\left({\Pd \ep _{B}\over\Pd 
\mlogg}\right)  +  \left({\Pd \ep _{A}\over\Pd \mlogg}\right)
\left({\Pd \ep _{B}\over\Pd T}\right)\biggr] \sigma_{T \mlogg} 
\nonumber\\ 
+ \biggl[\left({\Pd \ep _{A}\over\Pd \xi}\right)\left({\Pd \ep _{B}\over\Pd 
\mlogg}\right)  + 
\left({\Pd \ep _{A}\over\Pd \mlogg}\right)
\left({\Pd \ep _{B}\over\Pd \xi}\right)\biggr]\sigma_{\xi \mlogg}    
\end{eqnarray}}

\section{RESULTS AND DISCUSSION}
\label{sec:abund}

\subsection{Trends vs.\ \teff} 

\begin{figure}
\centering
      \includegraphics[width=0.9\hsize]{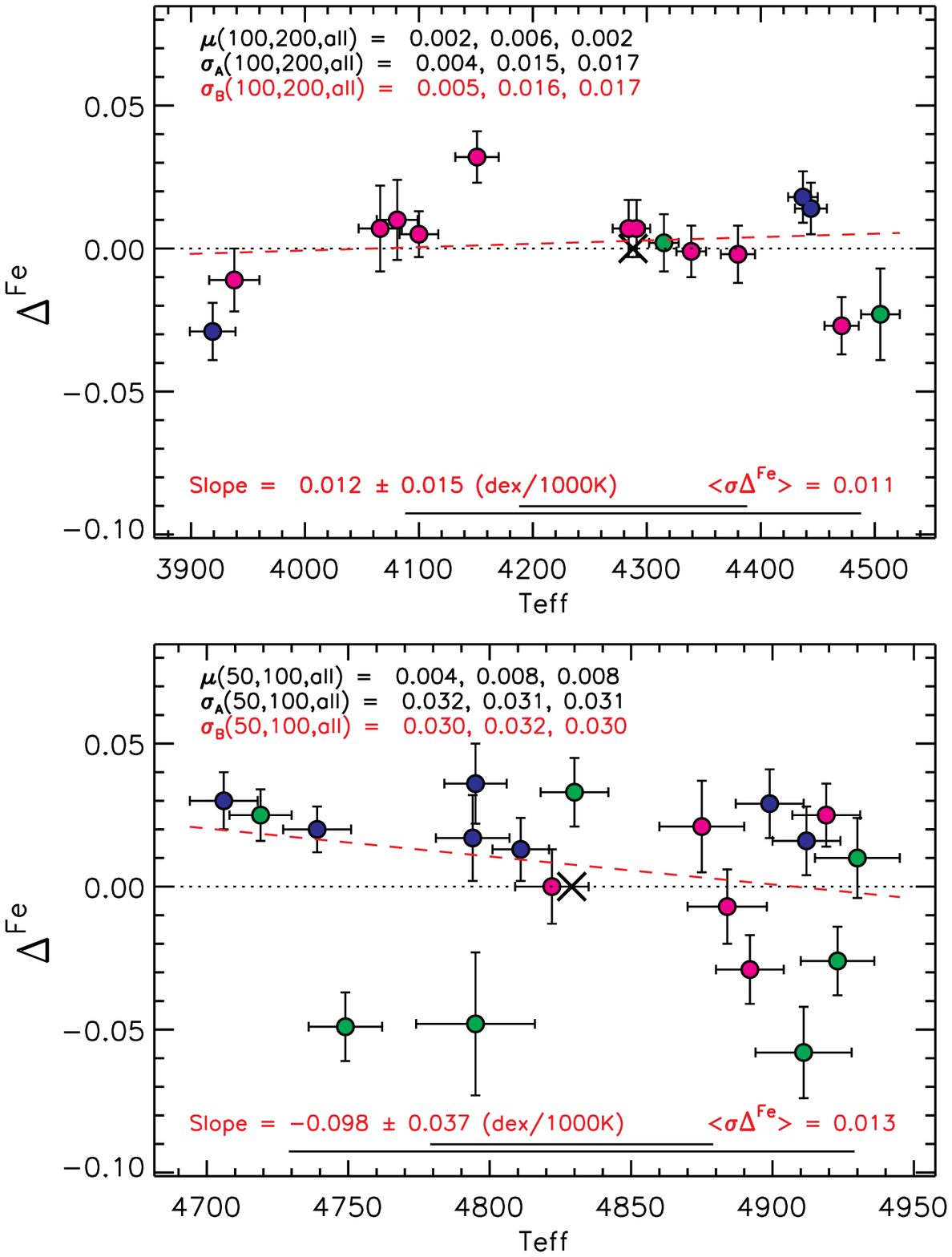} 
      \caption{$\Delta^{\rm Fe}$ vs.\ \teff\ for the RGB tip star sample (upper
panel) and the RGB bump star sample (lower panel). In both panels, we show the
location of the ``reference star'' as a black cross. We write the mean
abundance and standard deviation ($\sigma_{\rm A}$) for stars within 100K and
200K of the reference star as well as for the full sample. The red dashed line
is the linear least squares fit to the data. The slope, uncertainty and
dispersion ($\sigma_{\rm B}$) about the linear fit are written. We also write
the average abundance error, $<\sigma\Delta^{\rm Fe}>$, for each sample. (These
results are obtained when using the reference stars RGB tip = NGC 6752-mg9 and
RGB bump = NGC 6752-11.) The colours are the same as in Figure
\ref{fig:paramcomptip}}
         \label{fig:feteff}
\end{figure}

\begin{figure}
\centering
      \includegraphics[width=0.9\hsize]{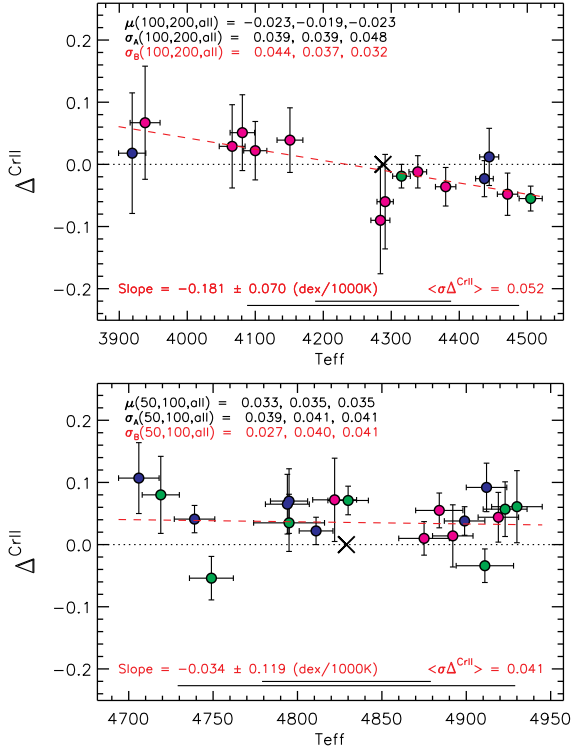} 
      \caption{Same as Figure \ref{fig:feteff} but for $\Delta^{\rm CrII}$ vs.\
\teff.}
         \label{fig:criiteff}
\end{figure}

\begin{figure}
\centering
      \includegraphics[width=0.9\hsize]{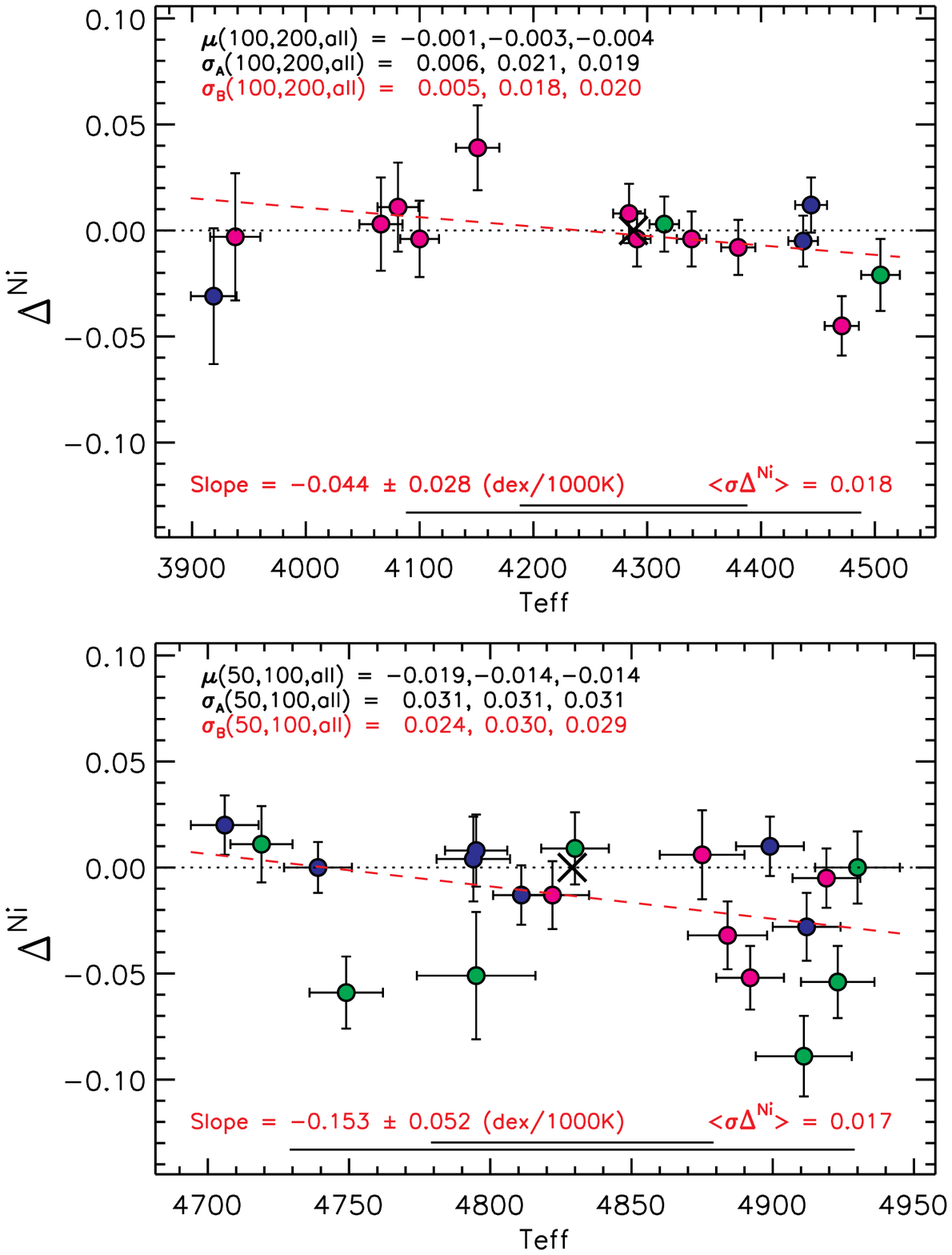} 
      \caption{Same as Figure \ref{fig:feteff} but for $\Delta^{\rm Ni}$ vs.\ \teff.}
         \label{fig:niteff}
\end{figure}

In Figures \ref{fig:feteff}, \ref{fig:criiteff} and \ref{fig:niteff}, we plot
$\Delta^{\rm Fe}$, $\Delta^{\rm CrII}$ and $\Delta^{\rm Ni}$ versus \teff,
respectively. In these figures, the RGB tip sample and the RGB bump sample are
in the upper and lower panels, respectively. In each panel, we show the mean
and the abundance dispersion for $\Delta^{\rm X}$ ($\sigma_{\rm A}$ in these
figures). We also determine the linear least squares fit to the data and write
the slope, uncertainty and abundance dispersion about the fit ($\sigma_{\rm
B}$ in these figures). For the subset of RGB tip stars within 100 K and 200 K
of the reference star, we compute and write the mean abundance and abundance
dispersions ($\sigma_{\rm A}$ and $\sigma_{\rm B}$). Similarly, for the subset
of RGB bump stars within 50 K and 100 K of the reference star, we write the
same quantities. Finally, we also write the average abundance error,
$<\sigma\Delta^{\rm X}>$, for a particular element for the RGB tip and RGB bump
samples. 

\begin{figure}
\centering
      \includegraphics[width=0.99\hsize]{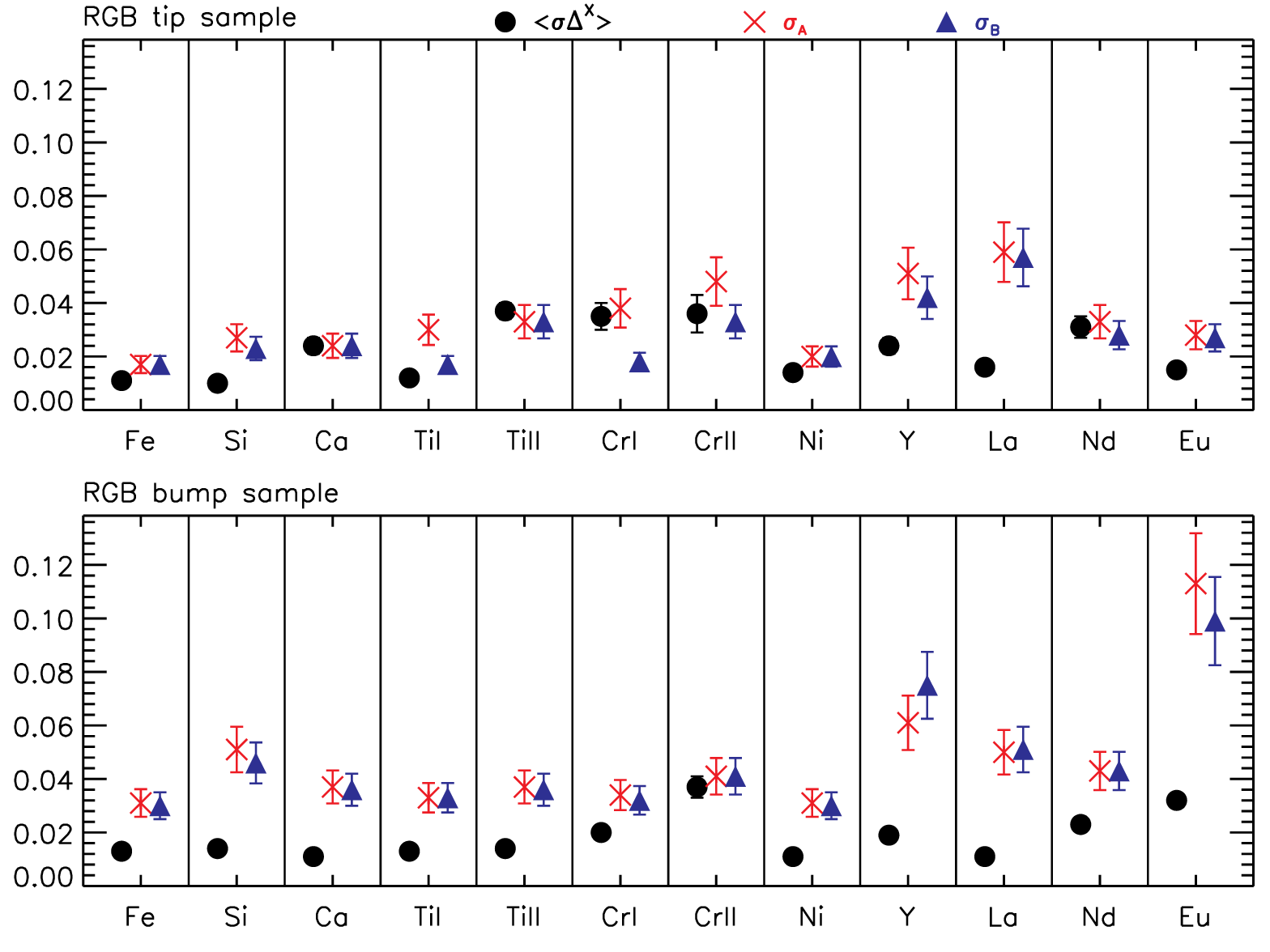} 
      \caption{Average abundance errors ($<\sigma\Delta^{X}>$, filled black circles),
abundance dispersions ($\sigma_{\rm A}$, red crosses) and abundance
dispersions about the linear fits as seen in Figures \ref{fig:feteff} to
\ref{fig:niteff} ($\sigma_{\rm B}$, blue triangles) for all species in the RGB
tip sample (upper panel) and RGB bump sample (lower panel). (These results are
obtained when using the reference stars RGB tip = NGC 6752-mg9 and RGB bump =
NGC 6752-11.)}
         \label{fig:abunerr}
\end{figure}

Fe and Ni (Figures \ref{fig:feteff} and \ref{fig:niteff}) are examples where
the average abundance errors are very small, $\sim$0.01 dex. \crii\ (Figure
\ref{fig:criiteff}) is the element that shows the highest average abundance
error, $\sim$0.04 dex. Rather than showing similar figures for every element,
in Figure \ref{fig:abunerr} we plot ($i$) the average abundance error
($<\sigma\Delta^{\rm X}>$), ($ii$) the abundance dispersion ($\sigma_{\rm A}$) 
and ($iii$) the abundance dispersion about the linear fit to $\Delta^{\rm X}$
versus \teff\ ($\sigma_{\rm B}$), for all elements in the RGB tip sample
(upper) and the RGB bump sample (lower). The main point to take from this
figure is that we have achieved very high precision chemical abundance
measurements from our strictly differential analysis for this sample of giant
stars in the globular cluster NGC 6752. For the RGB tip sample, the lowest
average abundance error is for Fe ($<\sigma\Delta^{\rm Fe}>$ = 0.011 dex) and
the highest value is for CrII ($<\sigma\Delta^{\rm CrII}>$ = 0.052 dex). For
the RGB bump sample the lowest average abundance errors are for Fe and La
($<\sigma\Delta^{\rm Fe,La}>$ = 0.013 dex) while the highest value is for CrII
($<\sigma\Delta^{\rm CrII}>$ = 0.041 dex).  Another aspect to note in Figure
\ref{fig:abunerr} is that the measured dispersions ($\sigma_{\rm A}$ and
$\sigma_{\rm B}$) for many elements appear to be considerably larger than the
average abundance error. We interpret such a result as evidence for a genuine
abundance dispersion in this cluster, although another possible explanation is
that we have systematically underestimated the errors. 

\subsection{$\Delta^{\rm X}$ vs.\ $\Delta^{\rm Na}$} 

\begin{figure}
\centering
      \includegraphics[width=0.9\hsize]{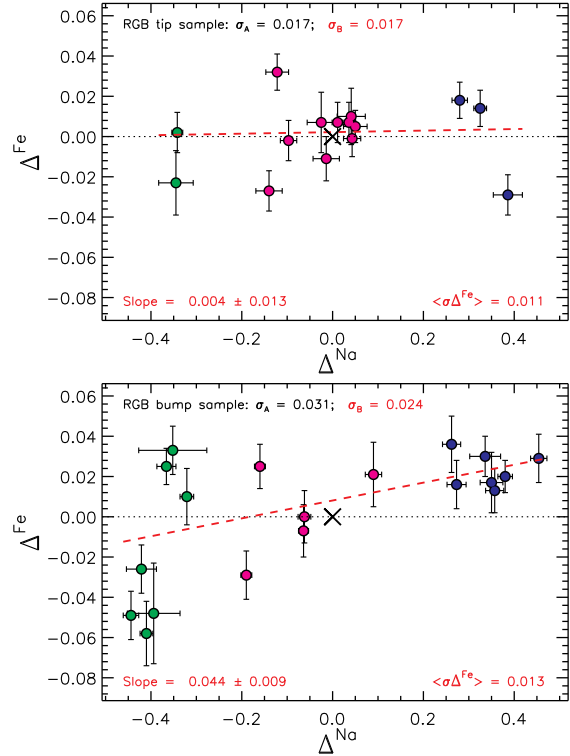} 
      \caption{$\Delta^{\rm Fe}$ vs.\ $\Delta^{\rm Na}$ for the RGB tip star sample
(upper) and the RGB bump star sample (lower). The red dashed line is the linear
least squares fit to the data (slope and error are written). We write the
dispersion in the $y$-direction ($\sigma_{\rm A}$), the dispersion about the
linear fit ($\sigma_{\rm B}$) and the average abundance error,
$<\sigma\Delta^{\rm Fe}>$, for each sample. (These results are obtained when
using the reference stars RGB tip = NGC 6752-mg9 and RGB bump = NGC 6752-11.)
The colours are the same as in Figure \ref{fig:paramcomptip}.}
         \label{fig:fena}
\end{figure}

\begin{figure}
\centering
      \includegraphics[width=0.9\hsize]{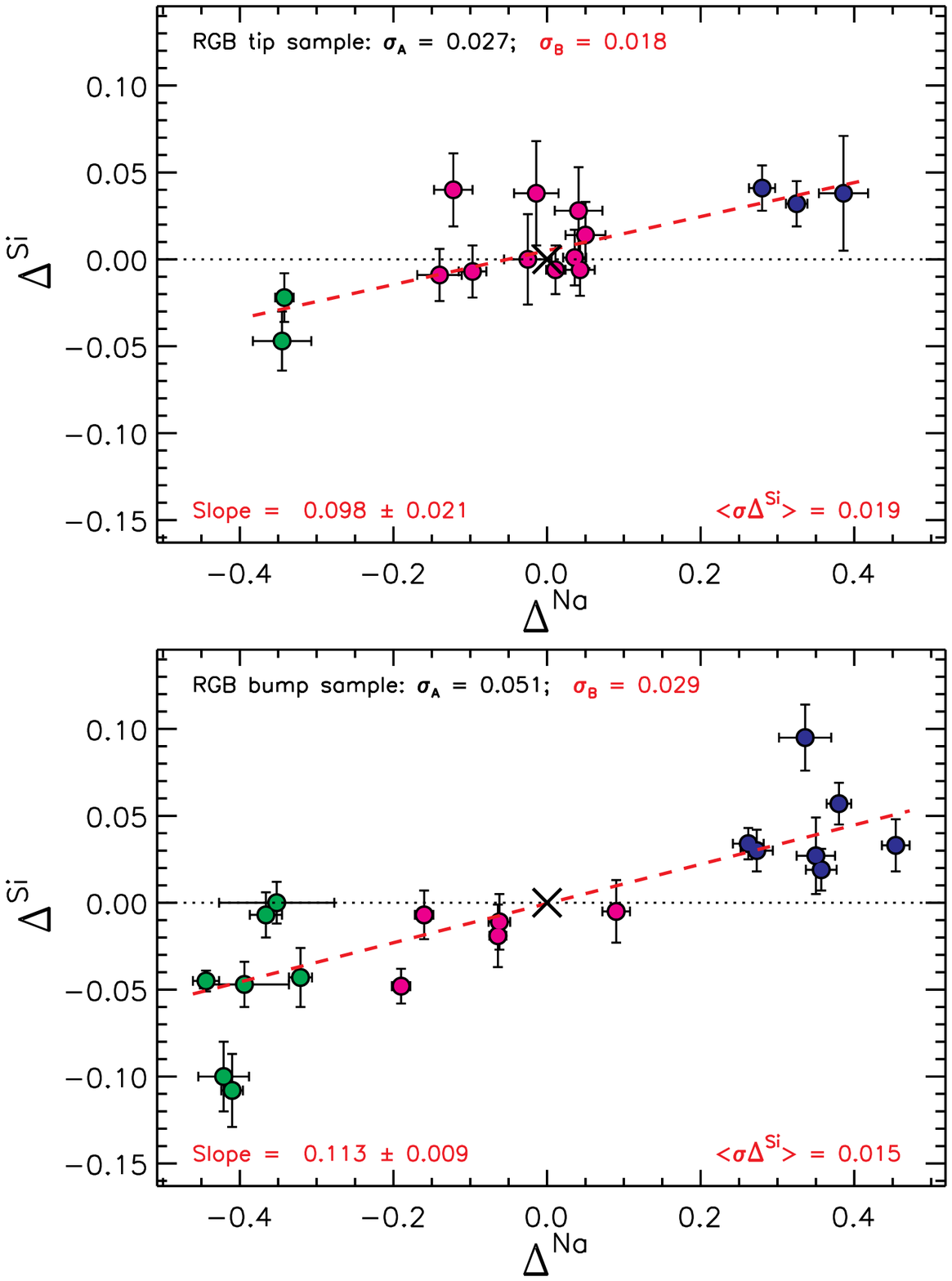} 
      \caption{Same as Figure \ref{fig:fena} but for $\Delta^{\rm Si}$ vs.\
$\Delta^{\rm Na}$.}
         \label{fig:sina}
\end{figure}

In Figures \ref{fig:fena} and \ref{fig:sina}, we plot $\Delta^{\rm Fe}$ vs.\
$\Delta^{\rm Na}$ and $\Delta^{\rm Si}$ vs.\ $\Delta^{\rm Na}$, respectively.
In both figures, the RGB tip sample and the RGB bump sample are found in the
upper and lower panels, respectively. (Here one readily sees that the
populations $a$ (green), $b$ (magenta) and $c$ (blue) identified by
\citet{milone13} from colour-magnitude diagrams have distinct Na abundances.)
We measure the linear least squares fit to the data and in each panel we write
($i$) the slope and uncertainty, ($ii$) the abundance dispersion ($\sigma_{\rm
A}$), ($iii$) the abundance dispersion about the linear fit to $\Delta^{\rm X}$
versus $\Delta^{\rm Na}$ ($\sigma_{\rm B}$) and (iv) the average abundance
error ($<\sigma\Delta^{\rm X}>$).  Consideration of the slope and uncertainty
of the linear fits reveals that while the amplitude may be small, there are
statistically significant correlations between $\Delta^{\rm Fe}$ and
$\Delta^{\rm Na}$ for the RGB bump sample and between $\Delta^{\rm Si}$ and
$\Delta^{\rm Na}$ for the RGB tip and RGB bump samples. The results for Si
confirm and expand on the correlations found between Si and Al \citep{yong05}
and between Si and N \citep{yong08nh}. 

\begin{figure}
\centering
      \includegraphics[width=0.9\hsize]{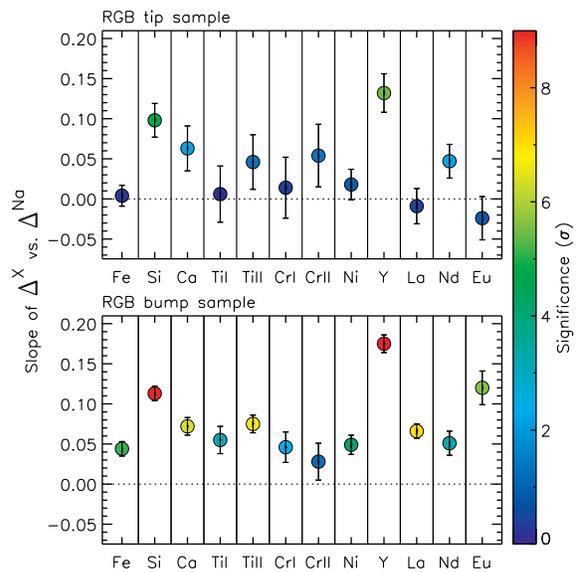} 
      \caption{Slope of the fit to $\Delta^{\rm X}$ vs.\ $\Delta^{\rm Na}$, for X =
Si to Eu, for the RGB tip sample (upper) and the RGB bump sample (lower). The
colours represent the significance of the slope, i.e., the magnitude of the
gradient divided by the uncertainty. (These results are obtained when using the
reference stars RGB tip = NGC 6752-mg9 and RGB bump = NGC 6752-11.)}
         \label{fig:allna}
\end{figure}

In Figure \ref{fig:allna}, we plot the slope of the linear fit to $\Delta^{\rm
X}$ vs.\ $\Delta^{\rm Na}$ for all elements in the RGB tip sample (upper) and
the RGB bump sample (lower). With the exception of La and Eu (for the RGB tip
sample), all the gradients are positive. For La and Eu in the RGB tip sample,
the negative gradients are not statistically significant, $<$1$\sigma$.
Assuming an equal likelihood of obtaining a positive or negative gradient, the
probability of obtaining 22 positive values in a sample of 24 is
$\sim$10$^{-5}$.  Based on the slope and uncertainty, we obtain the
significance of the correlations; eight of the 24 elements exhibit correlations
that are significant at the 5$\sigma$ level or higher\footnote{We also
performed linear fits to these data using the GaussFit program for robust
estimation \citep{jefferys88}. While we again find positive gradients for 22 of
the 24 elements, on average the significance of these correlations decreases
from 3.9$\sigma$ (least squares fitting) to 2.6$\sigma$ (robust fitting). When
using the GaussFit robust fitting routines, three of the 24 elements exhibit
correlations that are significant at the 5$\sigma$ level or higher.}.
Therefore, {\it the first main conclusion we draw is that there are an
unusually large number of elements that show positive correlations for
$\Delta^{\rm X}$ vs.\ $\Delta^{\rm Na}$, and that an unusually large fraction
of these correlations are of high statistical significance}. We interpret this
result as further evidence for a genuine abundance dispersion in this cluster.
On this occasion, it is highly unlikely that such correlations could arise from
underestimating the errors. NLTE corrections for Na, using improved atomic
data, have been published by \citet{lind11na}. The corrections are negative and
strongly dependent on line strength; for a given \teff:\logg:[Fe/H], stronger
lines have larger amplitude (negative) NLTE corrections. Had we included these
corrections, the $\Delta^{\rm X}$ vs.\ $\Delta^{\rm Na~(NLTE)}$ gradients would
be even steeper. 

We also note that the gradients are, on average, of larger amplitude and of
higher statistical significance for the RGB bump sample compared to the RGB tip
sample. Other than spanning a different range in stellar parameters, one
notable difference between the two samples is that the RGB bump sample exhibits
a larger range in $\Delta^{\rm Na}$ than does the RGB tip sample. In
particular, the numbers of RGB tip and RGB bump stars with $|\Delta^{\rm Na}|$
$\ge$ 0.20 dex are five and 14, respectively. (Equivalently, the numbers of
stars in the \citet{milone13} $b$ and $c$ populations are considerably larger
in the RGB bump sample compared to the RGB tip sample.) Thus, we speculate that
the RGB bump stars are the more reliable sample (based on the sample size and
abundance distribution) from which to infer the presence of any trend between
$\Delta^{\rm X}$ vs.\ $\Delta^{\rm Na}$. 

We conducted the following test in order to check whether differences in
gradients for $\Delta^{\rm X}$ vs.\ $\Delta^{\rm Na}$ between the RGB tip and
RGB bump samples can be attributed to differences in the Na distributions
between the two samples. We start by assuming that the RGB bump sample provides
the ``correct'' slope. For a given element, we consider the gradient and
uncertainty for $\Delta^{\rm X}$ vs.\ $\Delta^{\rm Na}$ and draw a random
number from a normal distribution (centered at zero) whose width corresponds to
the uncertainty.  We add that random number to the gradient to obtain a ``new
RGB bump gradient'' for $\Delta^{\rm X}$ vs.\ $\Delta^{\rm Na}$. For each RGB
tip star, we infer the corresponding $\Delta^{\rm X}$ using this ``new RGB bump
gradient''. We then draw another random number from a normal distribution
(centered at zero) of width corresponding to the measurement uncertainty,
$\sigma\Delta^{\rm X}$, and add that number to the $\Delta^{\rm X}$ value
inferred. For a given element, we measure the gradient and uncertainty for this
new set of $\Delta^{\rm X}$ values. We repeated the process for 1,000,000
realisations. Our expectation is that these Monte Carlo simulations predict the
gradient for $\Delta^{\rm X}$ vs.\ $\Delta^{\rm Na}$ that would be obtained
when combining ($i$) the RGB bump sample gradient with ($ii$) the RGB tip
sample Na distribution, and this approach accounts for the uncertainties in the
RGB bump sample gradients and measurement errors appropriate for the RGB tip
sample. For all elements except Fe (61123) and Eu (543)\footnote{The values in
parentheses refer to the numbers of realisations in which the gradient in the
simulations was consistent with the measured gradient. Fe is a $\sim$2$\sigma$
outlier. While Eu is clearly an outlier, we note that the abundances are
derived from a single line that is rather weak in the RGB bump stars.}, the
gradients measured from the RGB tip sample are consistent with those from the
simulations. We thus conclude that for most, but not all, elements the
differences in the $\Delta^{\rm X}$ vs.\ $\Delta^{\rm Na}$ gradients for the
two samples can be attributed to the differences in the Na distribution. 

\subsection{$\Delta^{\rm X}$ vs.\ $\Delta^{\rm Y}$} 

\begin{figure}
\centering
      \includegraphics[width=0.9\hsize]{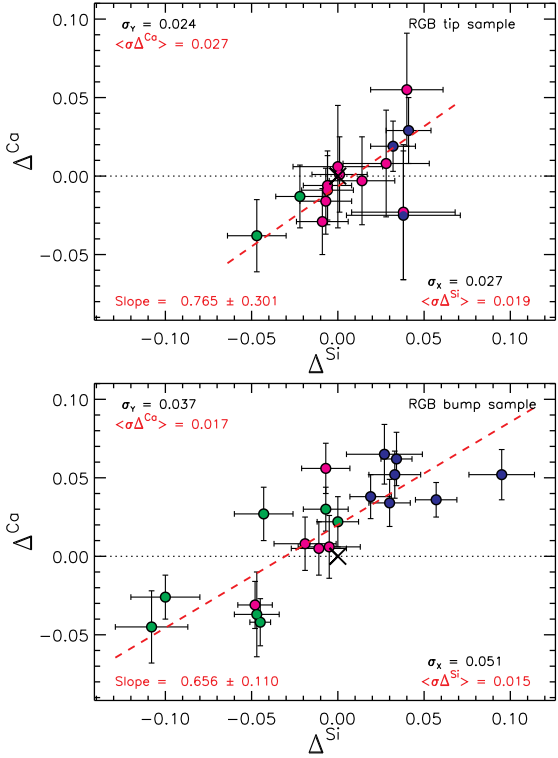} 
      \caption{$\Delta^{\rm Ca}$ vs.\ $\Delta^{\rm Si}$ for the RGB tip sample
(upper) and the RGB bump sample (lower). The red dashed line is the linear
least squares fit to the data (slope and error are written). We write the
abundance dispersions in the $x-$direction ($\sigma_{\rm X}$) and $y$-direction
($\sigma_{\rm Y}$) and the average abundance errors, $<\sigma\Delta^{\rm
Ca,Si}>$.  (These results are obtained when using the reference stars RGB tip =
NGC 6752-mg9 and RGB bump = NGC 6752-11.) The colours are the same as in Figure
\ref{fig:paramcomptip}.}
         \label{fig:sica}
\end{figure}

\begin{figure}
\centering
      \includegraphics[width=0.9\hsize]{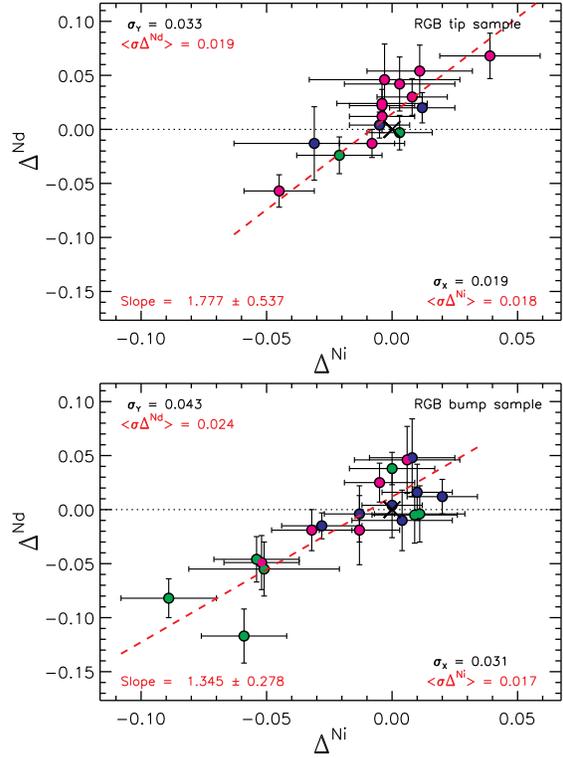} 
      \caption{Same as Figure \ref{fig:sica} but for $\Delta^{\rm Nd}$ vs.\
$\Delta^{\rm Ni}$.}
         \label{fig:nind}
\end{figure}

We now consider $\Delta^{\rm X}$ vs.\ $\Delta^{\rm Y}$, for every possible
combination of elements. In Figures \ref{fig:sica} and \ref{fig:nind} we plot
$\Delta^{\rm Ca}$ vs.\ $\Delta^{\rm Si}$ and $\Delta^{\rm Nd}$ vs.\
$\Delta^{\rm Ni}$, respectively. Once again we plot the linear least squares
fit to the data and write the slope and uncertainty. Consideration of those
quantities reveals that these pairs of elements show a statistically
significant correlation, although the amplitudes of the abundance variations
are small. In these figures, we write the abundance dispersions and average
abundance errors in the $x$-direction and the $y$-direction. As seen in Figure
\ref{fig:abunerr}, the abundance dispersions are almost always equal to, and in
some cases substantially larger than, the average measurement uncertainty. 

\begin{figure}
\centering
      \includegraphics[width=0.95\hsize]{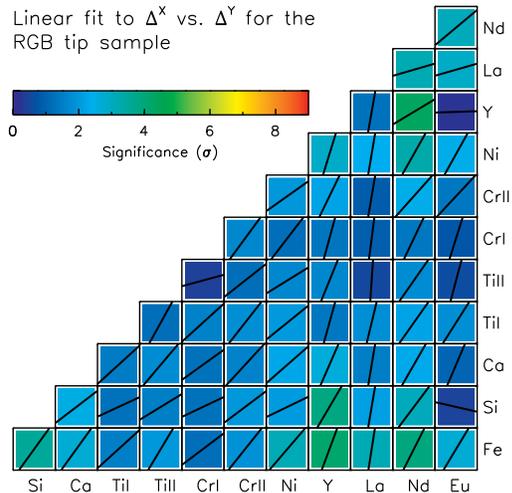} 
      \caption{Linear fit to $\Delta^{\rm X}$ vs.\ $\Delta^{\rm Y}$, for all
combination of elements, for the RGB tip sample. The dimensions of the x-axis
and y-axis are unity, such that a slope of gradient 1.0 would be represented by
a straight line from the lower left corner to the upper right corner and a
slope of gradient 0.0 would be a horizontal line. The significance of the
gradients are indicated by the colour bar. (These results are obtained when
using the reference stars RGB tip = NGC 6752-mg9 and RGB bump = NGC 6752-11.)}
         \label{fig:summarytip1}
\end{figure}

In Figure \ref{fig:summarytip1}, we show the linear fit to $\Delta^{\rm X}$
vs.\ $\Delta^{\rm Y}$ for all combinations of elements for the RGB tip sample.
The significance for a pair of elements, which is based on the slope and the
uncertainty, is shown in this figure. The gradients are always positive, with
the exception of the following pair of elements, Si and Eu (consideration of
the uncertainty suggests that the gradients are not significant). That is, 65
out of 66 pairs of elements exhibit a positive correlation\footnote{When using
the GaussFit robust estimation for the RGB tip sample, 64 out of 66 pairs of
elements exhibit a positive correlation. On average, the correlations for the
robust fitting (3.6$\sigma$) are of higher statistical significance than for
the least squares fitting (2.0$\sigma$) and 15 pairs of elements exhibit
correlations at the 5$\sigma$ level or higher. The average gradient is 2.06
$\pm$ 0.26 ($\sigma$ = 2.11) which is similar to the linear least squares
fitting.}.  The average gradient is 2.14 $\pm$ 0.29 ($\sigma$ = 2.37). 

\begin{figure}
\centering
      \includegraphics[width=0.95\hsize]{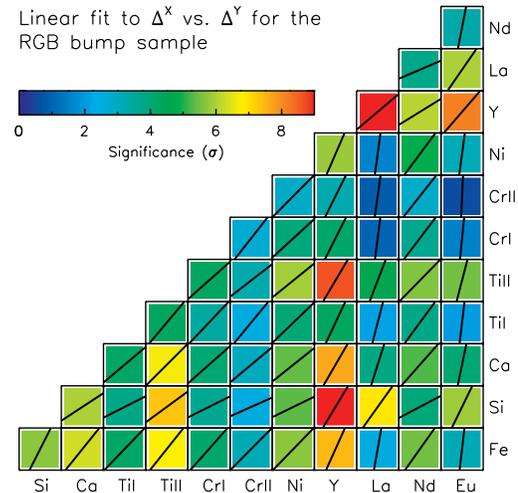} 
      \caption{Same as Figure \ref{fig:summarytip1} but for the RGB bump sample.}
         \label{fig:summarybum1}
\end{figure}

Figure \ref{fig:summarybum1} is the same as Figure \ref{fig:summarytip1}, but
for the RGB bump sample. The gradients are always positive with an average
value of 2.52 $\pm$ 0.40 ($\sigma$ = 3.29). Interestingly, the gradients are,
in general, of considerably higher statistical significance than in the RGB tip
sample. The average significance of the correlations is 2.0$\sigma$ for the
RGB tip sample and 4.5$\sigma$ for the RGB bump sample. For the RGB bump
sample, 25 pairs of elements (out of a total of 66) exhibit correlations
that are significant at the 5$\sigma$ level or higher\footnote{When using the
GaussFit robust estimation for the RGB bump sample, all pairs of elements
exhibit positive gradients. On average, the correlations for the robust fitting
(5.9$\sigma$) are of higher statistical significance than for the least squares
fitting (4.0$\sigma$) and 36 pairs of elements exhibit correlations at the
5$\sigma$ level or higher. The average gradient is 3.04 $\pm$ 0.65 ($\sigma$ =
5.30) and is only slightly higher than for the linear least squares fitting.}.
Thus, {\it the second main conclusion we draw is that there are an unusually
large number of elements that show positive correlations for $\Delta^{\rm X}$
vs.\ $\Delta^{\rm Y}$ and that many of these pairs of elements exhibit
correlations that are of high statistical significance}. Again, we speculate
that the higher statistical significance for the correlations between pairs of
elements in the RGB bump sample, compared to the RGB tip sample, is due to the
sample size and abundance distribution (i.e., the RGB bump sample includes many
more stars at the extremes of the $\Delta^{\rm Na}$, and therefore $\Delta^{\rm
X}$, distributions). Monte Carlo simulations indicate that the gradients for
the RGB bump and RGB tip samples are consistent when taking into account the
different distributions in $\Delta^{\rm X}$ between the two samples. We
interpret the significant correlations bewteen $\Delta^{\rm X}$ and
$\Delta^{\rm Y}$ as further indication of a genuine abundance dispersion in
this globular cluster. 

\subsection{Removing Trends With \teff}

Inspection of Figures \ref{fig:feteff}, \ref{fig:criiteff} and \ref{fig:niteff}
suggests that there are statistically significant trends between $\Delta^{\rm
X}$ and \teff. We tentatively attribute those abundance trends with \teff\ to
differential non-LTE effects and/or 3D effects (e.g., \citealt{asplund05}). In
this subsection, we explore whether or not our results change if we remove the
abundance trends with \teff. That is, do the abundance trends between ($i$)
$\Delta^{\rm X}$ vs.\ $\Delta^{\rm Na}$ and ($ii$) $\Delta^{\rm X}$ vs.\
$\Delta^{\rm Y}$ persist, or disappear, if we remove the abundance trends with
\teff? 
 
\begin{figure}
\centering
      \includegraphics[width=0.9\hsize]{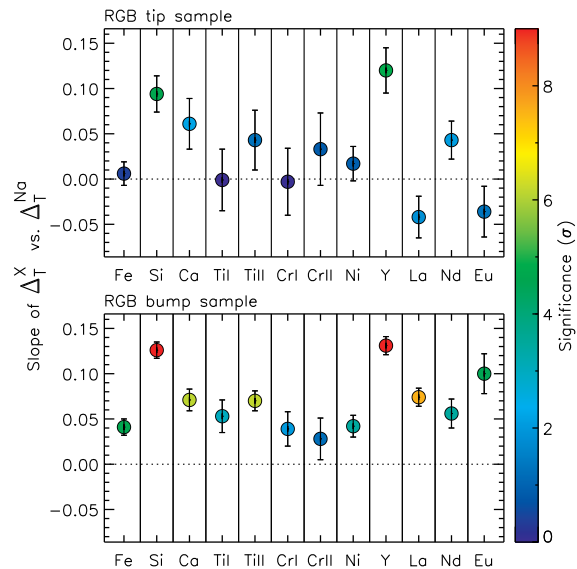} 
      \caption{Same as Figure \ref{fig:allna} but for $\Delta^{\rm X}_{\rm T}$ vs.\
$\Delta^{\rm Na}_{\rm T}$, i.e., the abundance trends with \teff\ have been
removed as described in Section 3.4. (These results are obtained when using the
reference stars RGB tip = NGC 6752-mg9 and RGB bump = NGC 6752-11.)}
         \label{fig:allnat}
\end{figure}

We remove those abundance trends with \teff\ in the following manner. We define
a new quantity, $\Delta^{\rm X}_{\rm T}$, as the difference between
$\Delta^{\rm X}$ and the value of the linear fit to the data at the \teff\ of
the program star.  In Figure \ref{fig:allnat}, we plot the slope of
$\Delta^{\rm X}_{\rm T}$ vs.\ $\Delta^{\rm Na}_{\rm T}$. This figure is similar
to Figure \ref{fig:allna}, but we have removed the abundance trends with \teff.
With the exception of Y in the RGB bump sample, our
results are unchanged at the $<$1.0$\sigma$ level. For Y, the slope and error
changed from 0.174 $\pm$ 0.011 to 0.131 $\pm$ 0.010, a difference of
3$\sigma$; in both cases the correlation is of high statistical significance. 

\begin{figure}
\centering
      \includegraphics[width=0.95\hsize]{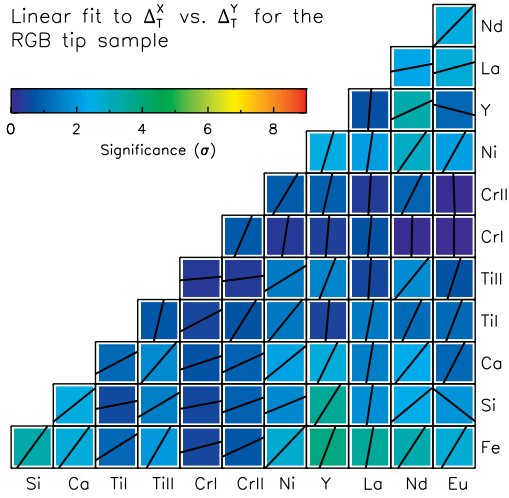} 
      \caption{Same as Figure \ref{fig:summarytip1} but for $\Delta^{\rm X}_{\rm T}$
vs.\ $\Delta^{\rm Y}_{\rm T}$ in the RGB tip sample, i.e., the abundance trends
with \teff\ have been removed as described in Section 3.4. (These results are
obtained when using the reference stars RGB tip = NGC 6752-mg9 and RGB bump =
NGC 6752-11.)}
         \label{fig:summarytip2}
\end{figure}

\begin{figure}
\centering
      \includegraphics[width=0.95\hsize]{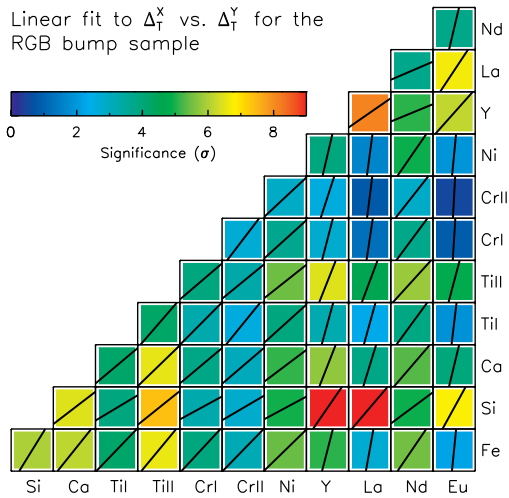} 
      \caption{Same as Figure \ref{fig:summarytip2} but for the RGB bump sample.}
         \label{fig:summarybum2}
\end{figure}

Next, we examine the trends between $\Delta^{\rm X}_{\rm T}$ vs.\ $\Delta^{\rm
Y}_{\rm T}$ (see Figures \ref{fig:summarytip2} and \ref{fig:summarybum2}).
These figures are the same as Figures \ref{fig:summarytip1} and
\ref{fig:summarybum1} but we have removed the abundance trends with \teff. On
comparing the RGB tip samples (Figures \ref{fig:summarytip1} vs.\
\ref{fig:summarytip2}) and the RGB bump samples (Figures \ref{fig:summarybum1}
vs.\ \ref{fig:summarybum2}), the results are unchanged, at the $<$2$\sigma$ 
level, for all pairs of elements. Therefore, we find positive correlations of
high statistical significance between pairs of elements regardless of whether
or not we remove any abundance trends with \teff. Such a result increases our
confidence that the abundance trends we identify are real and not an artefact
of systematic errors in the analysis. 

\subsection{Confirmation of Results When Using a Different Reference Star} 

An important consideration is whether or not the results change for a different
choice of reference stars. In this subsection, we repeat the entire analysis
but using a new pair of reference stars. For the RGB tip sample and RGB bump
sample, we use NGC 6752-mg6 and NGC 6752-1 as the reference stars,
respectively. These stars were arbitrarily chosen to have higher S/N (and
therefore lower \teff) than the previous pair of reference stars. 

Starting with the reference star parameters as described in Section 2.3, we 
obtained for each star in each sample, strictly differential stellar parameters
using the line-by-line analysis described in Section 2.4. The new strictly
differential stellar parameters are presented in Table \ref{tab:param2}. As
before, the strictly differential stellar parameters are very close to the
``reference star'' stellar parameters. 

\begin{table*}
 \centering
 \begin{minipage}{140mm}
  \caption{Strictly Differential Stellar Parameters, and Uncertainties, When
Adopting the Second Set of Reference Stars (RGB Tip = NGC 6752-mg6, RGB Bump =
NGC 6752-1). \label{tab:param2}} 
  \begin{tabular}{@{}lccccccc@{}}
  \hline
         Name & 
         \teff & 
         $\sigma$ & 
         \logg & 
         $\sigma$ & 
         \vt & 
         $\sigma$ & 
         [Fe/H] \\ 
          & 
         (K) & 
         (K) & 
         (cm s$^{-2}$) & 
         (cm s$^{-2}$) & 
         (\kms) & 
         (\kms) & 
          \\
         (1) & 
         (2) &
         (3) &
         (4) & 
         (5) & 
         (6) & 
         (7) & 
         (8) \\ 
\hline 
NGC6752-mg0  & 3922 & 20 & 0.19 & 0.01 & 2.24 & 0.04 & $-$1.68 \\
NGC6752-mg2  & 3940 & 16 & 0.25 & 0.01 & 2.11 & 0.04 & $-$1.66 \\
NGC6752-mg3  & 4070 & 14 & 0.55 & 0.01 & 1.92 & 0.03 & $-$1.64 \\
NGC6752-mg4  & 4087 & 14 & 0.57 & 0.01 & 1.90 & 0.03 & $-$1.64 \\
NGC6752-mg5  & 4105 & 16 & 0.59 & 0.01 & 1.93 & 0.04 & $-$1.64 \\
NGC6752-mg8  & 4288 & 17 & 0.98 & 0.01 & 1.71 & 0.04 & $-$1.64 \\
NGC6752-mg9  & 4292 & 20 & 0.96 & 0.01 & 1.73 & 0.05 & $-$1.65 \\
NGC6752-mg10 & 4295 & 14 & 0.96 & 0.01 & 1.69 & 0.04 & $-$1.64 \\
NGC6752-mg12 & 4315 & 17 & 1.00 & 0.01 & 1.73 & 0.05 & $-$1.65 \\
NGC6752-mg15 & 4347 & 17 & 1.04 & 0.01 & 1.77 & 0.05 & $-$1.65 \\
NGC6752-mg18 & 4387 & 13 & 1.10 & 0.01 & 1.70 & 0.04 & $-$1.65 \\
NGC6752-mg21 & 4443 & 16 & 1.19 & 0.01 & 1.69 & 0.06 & $-$1.63 \\
NGC6752-mg22 & 4451 & 18 & 1.23 & 0.01 & 1.71 & 0.07 & $-$1.64 \\
NGC6752-mg24 & 4511 & 16 & 1.33 & 0.01 & 1.70 & 0.06 & $-$1.67 \\
NGC6752-mg25 & 4479 & 15 & 1.28 & 0.01 & 1.72 & 0.06 & $-$1.67 \\
NGC6752-0    & 4737 & 11 & 1.86 & 0.01 & 1.44 & 0.02 & $-$1.62 \\
NGC6752-2    & 4770 & 10 & 1.95 & 0.01 & 1.36 & 0.02 & $-$1.63 \\
NGC6752-3    & 4781 & 11 & 1.98 & 0.01 & 1.36 & 0.02 & $-$1.70 \\
NGC6752-4    & 4827 & 12 & 2.07 & 0.01 & 1.39 & 0.02 & $-$1.63 \\
NGC6752-6    & 4830 & 13 & 2.10 & 0.01 & 1.34 & 0.02 & $-$1.61 \\
NGC6752-8    & 4966 & 16 & 2.29 & 0.01 & 1.33 & 0.03 & $-$1.64 \\
NGC6752-9    & 4829 & 18 & 2.08 & 0.01 & 1.42 & 0.03 & $-$1.69 \\
NGC6752-10   & 4846 & 12 & 2.10 & 0.01 & 1.38 & 0.02 & $-$1.63 \\
NGC6752-11   & 4866 &  6 & 2.13 & 0.01 & 1.37 & 0.02 & $-$1.64 \\
NGC6752-12   & 4855 & 13 & 2.14 & 0.01 & 1.35 & 0.02 & $-$1.64 \\
NGC6752-15   & 4866 & 15 & 2.23 & 0.01 & 1.37 & 0.02 & $-$1.61 \\
NGC6752-16   & 4911 & 15 & 2.24 & 0.01 & 1.33 & 0.03 & $-$1.62 \\
NGC6752-19   & 4928 & 12 & 2.32 & 0.01 & 1.33 & 0.02 & $-$1.67 \\
NGC6752-20   & 4935 & 13 & 2.33 & 0.01 & 1.32 & 0.02 & $-$1.62 \\
NGC6752-21   & 4921 & 14 & 2.32 & 0.01 & 1.32 & 0.03 & $-$1.65 \\
NGC6752-23   & 4945 & 12 & 2.32 & 0.01 & 1.26 & 0.02 & $-$1.63 \\
NGC6752-24   & 4945 & 14 & 2.39 & 0.01 & 1.15 & 0.03 & $-$1.70 \\
NGC6752-29   & 4959 & 12 & 2.40 & 0.01 & 1.32 & 0.02 & $-$1.67 \\
NGC6752-30   & 4954 & 13 & 2.47 & 0.01 & 1.25 & 0.02 & $-$1.62 \\
\hline
\end{tabular}
\end{minipage}
\end{table*}

With these revised stellar parameters, we computed chemical abundances and
conducted a full error analysis following the procedures outlined in Sections
2.5 and 2.6, respectively. In Tables \ref{tab:abun2a} and \ref{tab:abun2b} we
present the abundance differences for each element in all program stars when
using this new pair of reference stars.  (We did not, however, recompute
abundances based on spectrum synthesis analysis for La and Eu, and thus those
elements will not be considered in this subsection).  Again, we achieve high
precision chemical abundance measurements and the measured dispersions
($\sigma_{\rm A}$, $\sigma_{\rm B}$) are, in general, larger than the average
abundance error (particularly for the RGB bump sample). 

\begin{table*}
 \centering
 \begin{minipage}{160mm}
  \caption{Differential Abundances (Fe, Na, Si, Ca and Ti) When Adopting the
Second Set of Reference Stars (RGB Tip = NGC 6752-mg6, RGB Bump = NGC
6752-1).\label{tab:abun2a}} 
  \begin{tabular}{@{}lrcrcrcrcrcrc@{}}
  \hline
Star & 
$\Delta^{\rm Fe}$ &
$\sigma$ & 
$\Delta^{\rm Na}$ &
$\sigma$ & 
$\Delta^{\rm Si}$ &
$\sigma$ & 
$\Delta^{\rm Ca}$ &
$\sigma$ & 
$\Delta^{\rm TiI}$ &
$\sigma$ & 
$\Delta^{\rm TiII}$ &
$\sigma$ \\ 
(1) & 
(2) &
(3) &
(4) & 
(5) & 
(6) & 
(7) & 
(8) & 
(9) &
(10) & 
(11) & 
(12) & 
(13) \\ 
\hline
NGC6752-mg0  & $-$0.065 & 0.009 &    0.387 & 0.030 &    0.043 & 0.032 & $-$0.022 & 0.040 &    0.023 & 0.050 & $-$0.013 & 0.049 \\
NGC6752-mg2  & $-$0.047 & 0.011 & $-$0.014 & 0.017 &    0.044 & 0.019 & $-$0.017 & 0.028 &    0.054 & 0.035 &    0.050 & 0.038 \\
NGC6752-mg3  & $-$0.027 & 0.014 & $-$0.026 & 0.026 &    0.012 & 0.025 &    0.001 & 0.031 &    0.024 & 0.044 &    0.056 & 0.046 \\
NGC6752-mg4  & $-$0.024 & 0.010 &    0.043 & 0.024 &    0.037 & 0.020 &    0.012 & 0.025 &    0.029 & 0.034 &    0.058 & 0.037 \\
NGC6752-mg5  & $-$0.029 & 0.008 &    0.052 & 0.021 &    0.023 & 0.020 &    0.002 & 0.030 &    0.010 & 0.039 &    0.054 & 0.046 \\
NGC6752-mg8  & $-$0.036 & 0.016 &    0.038 & 0.002 &    0.007 & 0.006 &    0.008 & 0.011 &    0.015 & 0.010 &    0.052 & 0.066 \\
NGC6752-mg9  & $-$0.036 & 0.016 &    0.002 & 0.018 &    0.006 & 0.019 & $-$0.001 & 0.025 &    0.005 & 0.027 &    0.013 & 0.023 \\
NGC6752-mg10 & $-$0.028 & 0.011 &    0.015 & 0.014 &    0.003 & 0.016 & $-$0.002 & 0.022 & $-$0.018 & 0.024 &    0.043 & 0.017 \\
NGC6752-mg12 & $-$0.038 & 0.013 & $-$0.347 & 0.024 & $-$0.020 & 0.017 & $-$0.020 & 0.028 &    0.004 & 0.043 &    0.004 & 0.026 \\
NGC6752-mg15 & $-$0.036 & 0.013 &    0.048 & 0.024 & $-$0.002 & 0.018 & $-$0.005 & 0.031 & $-$0.002 & 0.040 &    0.022 & 0.033 \\
NGC6752-mg18 & $-$0.036 & 0.009 & $-$0.093 & 0.017 & $-$0.000 & 0.014 & $-$0.011 & 0.020 & $-$0.012 & 0.027 &    0.061 & 0.032 \\
NGC6752-mg21 & $-$0.018 & 0.011 &    0.283 & 0.019 &    0.048 & 0.015 &    0.034 & 0.025 & $-$0.007 & 0.031 &    0.072 & 0.033 \\
NGC6752-mg22 & $-$0.021 & 0.013 &    0.326 & 0.015 &    0.035 & 0.016 &    0.020 & 0.022 & $-$0.005 & 0.033 &    0.025 & 0.033 \\
NGC6752-mg24 & $-$0.056 & 0.015 & $-$0.341 & 0.038 & $-$0.043 & 0.016 & $-$0.033 & 0.023 & $-$0.026 & 0.027 &    0.066 & 0.063 \\
NGC6752-mg25 & $-$0.059 & 0.009 & $-$0.135 & 0.026 & $-$0.002 & 0.014 & $-$0.022 & 0.019 & $-$0.037 & 0.024 & $-$0.001 & 0.038 \\
NGC6752-0    &    0.006 & 0.009 &    0.699 & 0.051 &    0.105 & 0.024 &    0.021 & 0.014 &    0.020 & 0.021 &    0.019 & 0.016 \\
NGC6752-2    & $-$0.004 & 0.012 &    0.750 & 0.016 &    0.065 & 0.020 &    0.010 & 0.016 & $-$0.005 & 0.020 & $-$0.000 & 0.012 \\
NGC6752-3    & $-$0.071 & 0.012 & $-$0.075 & 0.010 & $-$0.032 & 0.018 & $-$0.069 & 0.015 & $-$0.054 & 0.020 & $-$0.070 & 0.011 \\
NGC6752-4    & $-$0.002 & 0.015 &    0.726 & 0.044 &    0.041 & 0.015 &    0.043 & 0.018 &    0.008 & 0.027 &    0.002 & 0.012 \\
NGC6752-6    &    0.019 & 0.016 &    0.636 & 0.014 &    0.048 & 0.014 &    0.039 & 0.019 &    0.030 & 0.027 &    0.015 & 0.013 \\
NGC6752-8    & $-$0.012 & 0.015 &    0.045 & 0.014 & $-$0.032 & 0.014 & $-$0.002 & 0.015 &    0.026 & 0.024 & $-$0.019 & 0.017 \\
NGC6752-9    & $-$0.065 & 0.022 & $-$0.024 & 0.038 & $-$0.034 & 0.017 & $-$0.060 & 0.024 & $-$0.061 & 0.037 & $-$0.074 & 0.012 \\
NGC6752-10   & $-$0.006 & 0.014 &    0.730 & 0.014 &    0.032 & 0.014 &    0.015 & 0.017 &    0.007 & 0.025 & $-$0.002 & 0.013 \\
NGC6752-11   & $-$0.019 & 0.006 &    0.373 & 0.021 &    0.016 & 0.013 & $-$0.024 & 0.011 &    0.001 & 0.015 & $-$0.030 & 0.012 \\
NGC6752-12   & $-$0.018 & 0.014 &    0.306 & 0.016 &    0.003 & 0.017 & $-$0.020 & 0.016 & $-$0.022 & 0.026 & $-$0.001 & 0.016 \\
NGC6752-15   &    0.013 & 0.015 &    0.018 & 0.056 &    0.013 & 0.019 & $-$0.003 & 0.019 & $-$0.005 & 0.027 &    0.010 & 0.013 \\
NGC6752-16   &    0.001 & 0.014 &    0.461 & 0.035 &    0.009 & 0.024 & $-$0.017 & 0.017 &    0.001 & 0.025 & $-$0.023 & 0.016 \\
NGC6752-19   & $-$0.050 & 0.013 &    0.179 & 0.014 & $-$0.034 & 0.022 & $-$0.056 & 0.014 & $-$0.048 & 0.020 & $-$0.055 & 0.013 \\
NGC6752-20   &    0.009 & 0.012 &    0.822 & 0.036 &    0.045 & 0.018 &    0.024 & 0.016 &    0.019 & 0.023 &    0.006 & 0.013 \\
NGC6752-21   & $-$0.027 & 0.013 &    0.308 & 0.023 & $-$0.005 & 0.019 & $-$0.015 & 0.017 & $-$0.009 & 0.022 & $-$0.021 & 0.012 \\
NGC6752-23   & $-$0.006 & 0.012 &    0.641 & 0.009 &    0.045 & 0.018 &    0.005 & 0.015 & $-$0.006 & 0.023 & $-$0.012 & 0.017 \\
NGC6752-24   & $-$0.079 & 0.012 & $-$0.041 & 0.013 & $-$0.094 & 0.014 & $-$0.076 & 0.018 & $-$0.082 & 0.022 & $-$0.110 & 0.013 \\
NGC6752-29   & $-$0.048 & 0.012 & $-$0.052 & 0.013 & $-$0.088 & 0.018 & $-$0.054 & 0.014 & $-$0.066 & 0.029 & $-$0.077 & 0.011 \\
NGC6752-30   &    0.004 & 0.012 &    0.207 & 0.013 &    0.005 & 0.013 &    0.030 & 0.016 &    0.001 & 0.024 &    0.020 & 0.015 \\
\hline
\end{tabular}
\end{minipage}
\end{table*}

\begin{table*}
 \centering
 \begin{minipage}{160mm}
  \caption{Differential Abundances (Cr, Ni, Y and Nd) When Adopting the Second
Set of Reference Stars (RGB Tip = NGC 6752-mg6, RGB Bump = NGC
6752-1).\label{tab:abun2b}}
  \begin{tabular}{@{}lrcrcrcrcrc@{}}
  \hline
Star & 
$\Delta^{\rm CrI}$ &
$\sigma$ & 
$\Delta^{\rm CrII}$ &
$\sigma$ & 
$\Delta^{\rm Ni}$ &
$\sigma$ & 
$\Delta^{\rm Y}$ &
$\sigma$ & 
$\Delta^{\rm Nd}$ &
$\sigma$ \\ 
(1) & 
(2) &
(3) &
(4) & 
(5) & 
(6) & 
(7) & 
(8) & 
(9) &
(10) & 
(11) \\ 
\hline
NGC6752-mg0  &    0.011 & 0.067 &    0.028 & 0.093 & $-$0.023 & 0.030 &    0.034 & 0.038 &    0.003 & 0.033 \\
NGC6752-mg2  &    0.053 & 0.090 &    0.076 & 0.077 &    0.007 & 0.027 &    0.101 & 0.049 &    0.067 & 0.030 \\
NGC6752-mg3  &    0.044 & 0.055 &    0.053 & 0.059 &    0.013 & 0.018 &    0.092 & 0.025 &    0.062 & 0.021 \\
NGC6752-mg4  &    0.053 & 0.052 &    0.068 & 0.049 &    0.021 & 0.018 &    0.090 & 0.023 &    0.075 & 0.021 \\
NGC6752-mg5  &    0.034 & 0.045 &    0.038 & 0.049 &    0.008 & 0.018 &    0.025 & 0.039 &    0.049 & 0.020 \\
NGC6752-mg8  & $-$0.025 & 0.044 & $-$0.079 & 0.024 &    0.018 & 0.012 &    0.039 & 0.031 &    0.052 & 0.014 \\
NGC6752-mg9  &    0.002 & 0.036 &    0.014 & 0.086 &    0.008 & 0.017 &    0.019 & 0.016 &    0.020 & 0.019 \\
NGC6752-mg10 & $-$0.006 & 0.028 & $-$0.040 & 0.076 &    0.008 & 0.015 &    0.100 & 0.024 &    0.042 & 0.016 \\
NGC6752-mg12 & $-$0.008 & 0.032 & $-$0.010 & 0.034 &    0.006 & 0.017 &    0.005 & 0.017 &    0.019 & 0.019 \\
NGC6752-mg15 & $-$0.024 & 0.032 & $-$0.006 & 0.036 &    0.002 & 0.017 &    0.014 & 0.016 &    0.032 & 0.018 \\
NGC6752-mg18 & $-$0.022 & 0.025 & $-$0.021 & 0.030 &    0.002 & 0.012 &    0.032 & 0.018 &    0.009 & 0.012 \\
NGC6752-mg21 & $-$0.000 & 0.031 & $-$0.009 & 0.038 &    0.005 & 0.014 &    0.085 & 0.021 &    0.029 & 0.014 \\
NGC6752-mg22 & $-$0.014 & 0.048 &    0.019 & 0.053 &    0.016 & 0.016 &    0.062 & 0.029 &    0.030 & 0.016 \\
NGC6752-mg24 & $-$0.028 & 0.022 & $-$0.047 & 0.019 & $-$0.015 & 0.016 & $-$0.045 & 0.015 & $-$0.012 & 0.016 \\
NGC6752-mg25 & $-$0.019 & 0.027 & $-$0.033 & 0.029 & $-$0.033 & 0.013 & $-$0.016 & 0.026 & $-$0.026 & 0.014 \\
NGC6752-0    &    0.021 & 0.021 &    0.034 & 0.021 &    0.012 & 0.016 &    0.020 & 0.012 &    0.030 & 0.021 \\
NGC6752-2    & $-$0.028 & 0.025 & $-$0.038 & 0.063 & $-$0.012 & 0.016 & $-$0.041 & 0.044 &    0.002 & 0.012 \\
NGC6752-3    & $-$0.088 & 0.023 & $-$0.127 & 0.085 & $-$0.064 & 0.021 & $-$0.171 & 0.030 & $-$0.104 & 0.015 \\
NGC6752-4    & $-$0.018 & 0.029 & $-$0.011 & 0.018 & $-$0.001 & 0.020 & $-$0.008 & 0.045 & $-$0.004 & 0.017 \\
NGC6752-6    &    0.008 & 0.039 & $-$0.001 & 0.014 &    0.002 & 0.020 & $-$0.016 & 0.031 &    0.056 & 0.028 \\
NGC6752-8    & $-$0.019 & 0.029 & $-$0.016 & 0.024 & $-$0.008 & 0.021 & $-$0.051 & 0.011 &    0.043 & 0.027 \\
NGC6752-9    & $-$0.071 & 0.040 & $-$0.042 & 0.023 & $-$0.056 & 0.030 & $-$0.109 & 0.017 & $-$0.051 & 0.011 \\
NGC6752-10   & $-$0.007 & 0.029 & $-$0.057 & 0.078 & $-$0.019 & 0.018 & $-$0.008 & 0.022 & $-$0.001 & 0.015 \\
NGC6752-11   & $-$0.034 & 0.015 & $-$0.072 & 0.061 & $-$0.002 & 0.015 & $-$0.019 & 0.028 &    0.015 & 0.025 \\
NGC6752-12   & $-$0.027 & 0.030 &    0.002 & 0.013 & $-$0.020 & 0.020 & $-$0.118 & 0.036 & $-$0.008 & 0.010 \\
NGC6752-15   & $-$0.012 & 0.033 & $-$0.000 & 0.049 &    0.002 & 0.020 & $-$0.066 & 0.009 &    0.005 & 0.017 \\
NGC6752-16   & $-$0.020 & 0.026 & $-$0.060 & 0.066 &    0.004 & 0.024 & $-$0.068 & 0.031 &    0.060 & 0.034 \\
NGC6752-19   & $-$0.073 & 0.026 & $-$0.056 & 0.018 & $-$0.055 & 0.015 & $-$0.127 & 0.028 & $-$0.034 & 0.012 \\
NGC6752-20   & $-$0.013 & 0.028 & $-$0.034 & 0.073 &    0.003 & 0.020 & $-$0.008 & 0.019 &    0.026 & 0.012 \\
NGC6752-21   & $-$0.049 & 0.024 & $-$0.021 & 0.057 & $-$0.035 & 0.019 & $-$0.033 & 0.014 & $-$0.007 & 0.015 \\
NGC6752-23   & $-$0.031 & 0.033 &    0.025 & 0.031 & $-$0.033 & 0.018 & $-$0.010 & 0.029 &    0.005 & 0.026 \\
NGC6752-24   & $-$0.093 & 0.023 & $-$0.103 & 0.077 & $-$0.095 & 0.023 & $-$0.156 & 0.015 & $-$0.061 & 0.024 \\
NGC6752-29   & $-$0.075 & 0.024 & $-$0.023 & 0.021 & $-$0.061 & 0.019 & $-$0.104 & 0.007 & $-$0.041 & 0.025 \\
NGC6752-30   & $-$0.006 & 0.026 & $-$0.026 & 0.026 & $-$0.012 & 0.017 & $-$0.022 & 0.043 &    0.038 & 0.016 \\
\hline
\end{tabular}
\end{minipage}
\end{table*}

\begin{figure}
\centering
      \includegraphics[width=0.9\hsize]{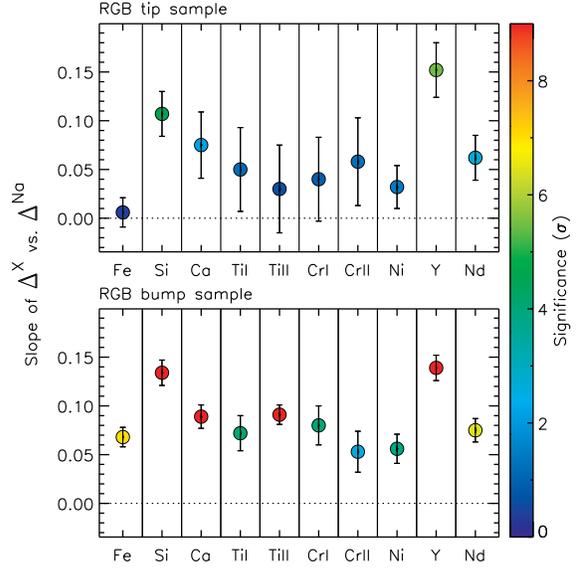} 
      \caption{Slope of the fit to $\Delta^{\rm X}$ vs.\ $\Delta^{\rm Na}$, for
X = Si to Eu, for the RGB tip sample (upper) and the RGB bump sample (lower).
The colours represent the significance of the slope. (This shows the same
results as Figure \ref{fig:allna} but for a different pair of reference stars,
RGB tip = NGC 6752-mg6 and RGB bump = NGC 6752-1.)} 
         \label{fig:allna2}
\end{figure}

\begin{figure}
\centering
      \includegraphics[width=0.9\hsize]{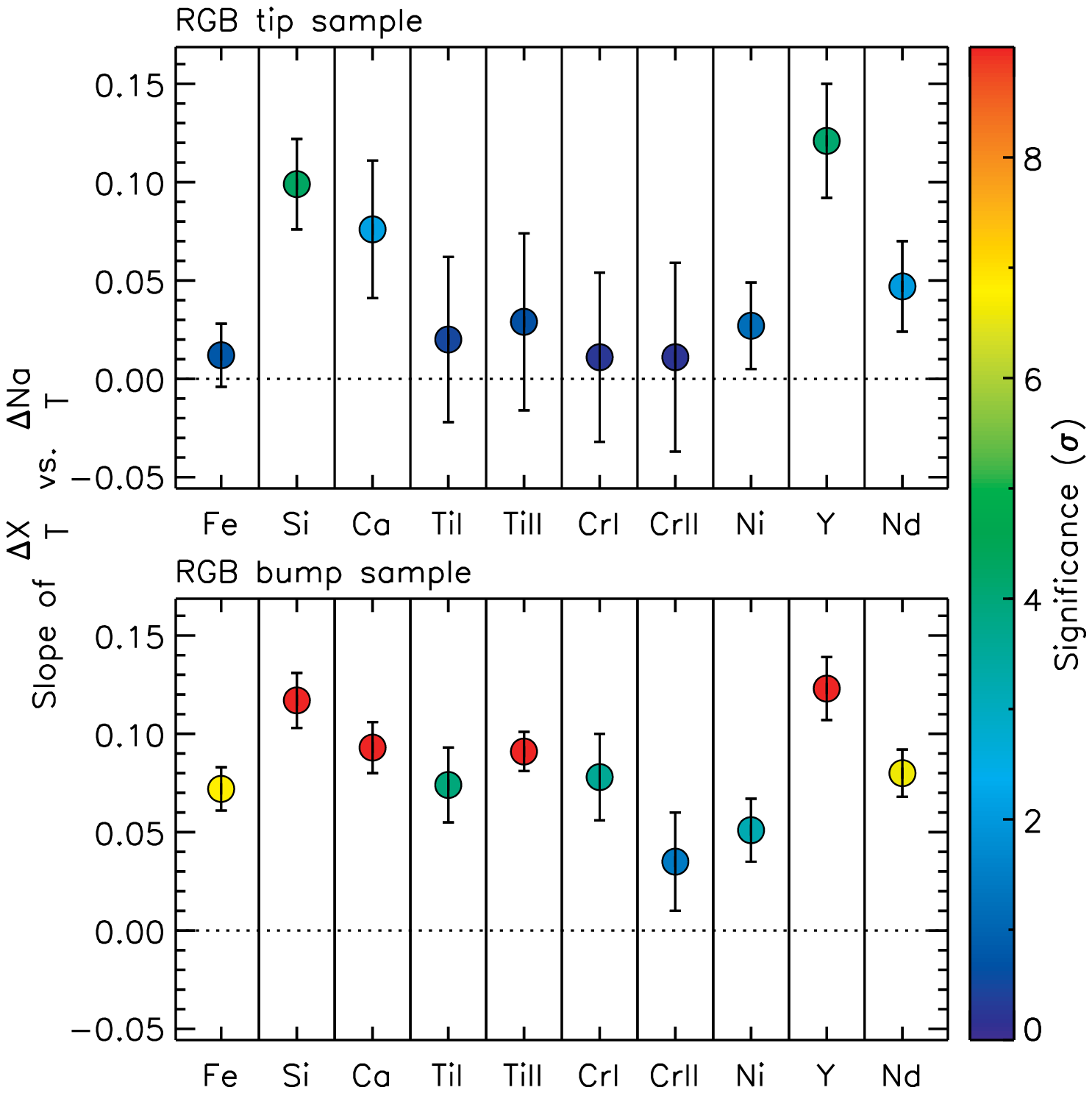} 
      \caption{Same as Figure \ref{fig:allna2} but for $\Delta^{\rm X}_{\rm T}$ vs.\
$\Delta^{\rm Na}_{\rm T}$, i.e., the abundance trends with \teff\ have been
removed as described in Section 3.4. (These results are obtained when using the
reference stars RGB tip = NGC 6752-mg6 and RGB bump = NGC 6752-1.)}
         \label{fig:allnat2}
\end{figure}

We examine the abundance trends $\Delta^{\rm X}$ versus $\Delta^{\rm Na}$ and
$\Delta^{\rm X}_{\rm T}$ versus $\Delta^{\rm Na}_{\rm T}$ in Figures 
\ref{fig:allna2} and \ref{fig:allnat2}, respectively.  As in Sections 3.2 and
3.4, we find that the abundance trends with Na are always positive and that a
large number of elements exhibit statistically significant correlations, albeit
of small amplitude. These results remain even after removing the abundance
trends as a function of \teff. 

Finally, we consider the abundance trends $\Delta^{\rm X}$ versus $\Delta^{\rm
Y}$. Our results are essentially identical to those in Sections 3.3 and 3.4,
namely, that for many pairs of elements, there are positive correlations of
high statistical significance for $\Delta^{\rm X}$ versus $\Delta^{\rm Y}$.
Again, these results remain even after removing the abundance trends with
\teff. 

The essential point to take from this subsection is that our results are not
sensitive to the choice of reference star, at least for the two cases we 
investigated. 

\subsection{Consequences for Globular Cluster Chemical Evolution} 

We begin with a summary of our analysis and results. 
\begin{enumerate}
\item From a strictly differential line-by-line analysis of a sample of RGB tip
stars and RGB bump stars in the globular cluster NGC 6752, we have obtained
revised stellar parameters which we refer to as ``strictly differential'' 
stellar parameters. 
\item Using those ``strictly differential'' stellar parameters, we have
computed differential chemical abundances, $\Delta^{\rm X}$ (for X = Fe, Na,
Si, Ca, Ti, Cr, Ni, Y, La, Nd and Eu), and conducted a detailed error
analysis. 
\item We have achieved very high precision measurements; for a given element,
our average relative abundance errors range from 0.01 dex to 0.05 dex. 
\item When plotting our abundance ratios against Na, e.g., $\Delta^{\rm X}$
versus $\Delta^{\rm Na}$, an unusually large number of elements show positive
correlations, often of high statistical significance, although the amplitudes
of the abundance variations in $\Delta^{\rm X}$ are small. 
\item When plotting the abundance ratios for any pair of elements, $\Delta^{\rm
X}$ versus $\Delta^{\rm Y}$, the majority exhibit positive correlations, often
of high statistical significance. 
\item Points (iv) and (v) persist even after ($a$) removing abundance trends
with \teff\ and/or ($b$) conducting a re-analysis using a different pair of
reference stars, thereby increasing our confidence in these results. 
\end{enumerate}
We now explore the consequences for globular cluster chemical evolution. 

At face value, our results would suggest that the globular cluster NGC 6752 is
not chemically homogeneous at the $\sim$0.03 dex level for the elements studied
here. Chemical inhomogeneity at this level can only be revealed when the
measurement uncertainties are $<$0.03 dex, as in this study. By extension, we
speculate that other globular clusters with no obvious dispersion in Fe-peak
elements but large Na variations (e.g., 47 Tuc, NGC 6397) may also display
similar behavior to NGC 6752 if subjected to a strictly differential chemical
abundance analysis of comparably high-quality spectra to that of this study. 

The abundance variations and positive correlations between $\Delta^{\rm X}$
versus $\Delta^{\rm Na}$ and between $\Delta^{\rm X}$ versus $\Delta^{\rm Y}$
could be due to a number of possibilities. Here we discuss four potential
scenarios, which are not mutually exclusive: 
(1) systematic errors in the stellar parameters; 
(2) star-to-star CNO abundance variations;
(3) star-to-star helium abundance variations; 
(4) inhomogeneous chemical evolution in the early stages of globular cluster
formation. 

\subsubsection{Systematic errors in the stellar parameters}

In the first scenario, we assume that the abundance variations are due to
systematic errors in the stellar parameters.  As noted in Section 3.1, the
abundance dispersions often exceed the average abundance error. Attributing the
abundance variations to systematic errors in the stellar parameters would
require a substantial underestimate of the stellar parameter uncertainties.
Such an explanation may be plausible. However, the abundance variations are
highly correlated and are seen for all elements which cover a variety of
ionization potentials and ionization states. There is no single change in
\teff, \logg\ or \vt\ that would remove the abundance correlations for all
elements in any given star. Thus, we regard this hypothesis to be unlikely. 

\subsubsection{Star-to-star CNO abundance variations}

In the second scenario, we assume that the abundance variations and
correlations are due to neglect of the appropriate C, N and O abundances in the
model atmospheres. The structure of the model atmosphere depends upon the
adopted C, N and O abundances \citep{gustafsson75}. \citet{drake93} studied the
effect of CNO abundances on the atmospheric structure in giant stars with
metallicities similar to that of NGC 6752. For the outer layers of the
atmosphere, the ``CN-weak'' models (i.e., appropriate for Na-poor objects) were
cooler than the ``CN-strong'' models (i.e., appropriate for Na-rich objects)
and the maximum difference was $\sim$ 150K.  The differences in abundances
derived using the ``CN-strong'' models minus those from the ``CN-weak'' models
for the \teff\ = 4400K, \logg\ = 1.3 and [Fe/H] = $-$1.5 case are almost all
positive and range from $\sim$ 0.00 dex to $\sim$ 0.10 dex. While the
magnitudes of the predicted abundance differences are similar to those of this
study, these differences have the incorrect sign.  That is, if we had analysed
the most Na-rich stars using the ``CN-strong'' models, according to the
\citet{drake93} predictions the inferred abundances would be higher and the
slope of the correlations between $\Delta^{\rm X}$ and $\Delta^{\rm Na}$ would
be even steeper.  We note, however, that the vast majority of our lines are
weak ($\log (W_\lambda/\lambda)$ $\le$ $-$5.0) such that the predicted
abundance differences are essentially zero and thus application of
``CN-strong'' models with appropriate CNO abundances to the Na-rich stars would
not change the trends we find. 

In the \citet{drake93} models, the C+N+O abundance sum was constant to within
0.12 dex between the ``CN-weak'' and ``CN-strong'' models. This assumption of
almost constant C+N+O abundance is appropriate for NGC 6752 on two grounds.
First, the presence of a substantial C+N+O abundance variation would manifest
as a spread in the luminosity of subgiant branch stars \citep{rood85} and such
a feature has not been detected in this cluster \citep{milone13}.  Second,
within their measurement uncertainties, \citet{carretta05} found no evidence
for a dispersion in the C+N+O abundance sum in NGC 6752 and preliminary work we
are conducting also indicates a nearly constant C+N+O abundance sum. 

\subsubsection{Star-to-star helium abundance variations}

In the third scenario, we assume that the abundance variations and correlations
are due to star-to-star He abundance variations. A
detailed analysis of the highest quality colour-magnitude diagrams available
shows that NGC 6752 harbours an internal He spread of up to $\Delta Y$ $\sim$
0.03 \citep{milone13}.  The most Na-rich objects are assumed to be more He-rich
relative to the Na-poor objects.  Spectroscopic analysis by \citet{villanova09}
showed that He measurements are possible in the cooler blue horizontal branch
stars of NGC 6752; they found a uniform He content, a result not unexpected
given the O-Na abundances of their targets. 

He abundance variations would affect our analysis in two distinct ways.  First,
the structure of the model atmosphere depends upon the adopted He abundance
\citep{stromgren82}.  Second, for a fixed mass fraction of metals ($Z$), a
change in the helium mass fraction ($Y$) will directly affect the hydrogen mass
fraction ($X$) such that the metal-to-hydrogen ratio, $Z$/$X$ will change with
helium mass fraction since $X+Y+Z=1$. We now consider both cases. 

Regarding the effect of He on the structure of a model atmosphere,  
\citet{stromgren82} demonstrated that for F type dwarfs, changes in the He/H
ratio ``affect the mean molecular weight of the gas and have an impact on the
gas pressure'' and that ``a helium-enriched atmosphere is similar to a
helium-normal atmosphere with a higher surface gravity, in terms of temperature
structure and electron pressure structure'' \citep{lind11}. Equation 12 in
\citet{stromgren82} quantifies the change in \logg\ due to a change in He/H
ratio; \citet{lind11} showed that metal-poor giants behave similarly. From this
equation, a change in He abundance from $Y$ = 0.25 to $Y$ = 0.28 would result
in a shift in \logg\ of 0.012. Inclusion of He abundance variations in the
model atmospheres would naively be expected to result in different stellar
parameters than those derived in this work, for both a regular analysis (as
used to define the reference star stellar parameters) and a strictly
differential analysis. Using a revised set of stellar parameters would, of
course, result in an updated set of chemical abundances (and line-by-line
chemical abundance differences). 

\begin{figure}
\centering
      \includegraphics[width=1.0\hsize]{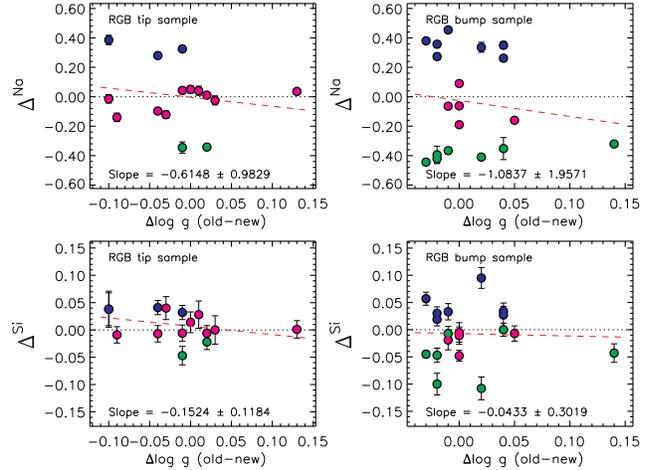} 
      \caption{$\Delta^{\rm Na}$ (upper) and $\Delta^{\rm Si}$ (lower) vs.\
$\Delta$\logg\ (old = ``reference star'' values, new = ``strictly
differential'' values) for the RGB tip sample (left) and the RGB bump sample
(right). The red dashed line is the linear fit to the data. (These results are
obtained when using the reference stars RGB tip = NGC 6752-mg9 and RGB bump =
NGC 6752-11.) As in Figure \ref{fig:paramcomptip}, the green, magenta and blue
colours represent populations $a$, $b$, and $c$, respectively, from
\citet{milone13} (see Section 2.1 for details).} 
         \label{fig:nadeltalogg}
\end{figure}

We might therefore expect to find a correlation between the Na abundance (which
is assumed to trace the He abundance) and the stellar parameters (or difference
between the strictly differential stellar parameters and the reference star
stellar parameters). In Figure \ref{fig:nadeltalogg}, we plot $\Delta^{\rm Na}$
against $\Delta$\logg\ (``reference star'' values minus ``strictly differential
analysis'' values).  There are no significant correlations for either the RGB
tip sample or the RGB bump sample. In light of the statistically significant
correlation between Si and Na, we also include in Figure \ref{fig:nadeltalogg}
panels showing $\Delta^{\rm Si}$ against $\Delta$\logg.  Again, there are no
significant correlations. (Similar plots using $\Delta$\teff\ rather than
$\Delta$\logg\ also reveal no significant correlations.) Given the magnitude of
the change in \logg\ resulting from the difference in helium abundance, it is
not surprising that we do not detect any significant trend between $\Delta^{\rm
Na}$ and $\Delta$\logg. Indeed, \citet{lind11} find that changes in helium of
$\Delta Y$ = 0.03, as is the case for NGC 6752, would be expected to result in
negligible changes in \teff\ and \logg. 

On the other hand, for a fixed mass fraction of metals ($Z$), a change in
the helium mass fraction ($Y$) will change the hydrogen mass fraction ($X$) and
the metal-to-hydrogen ratio, $Z$/$X$, since $X+Y+Z=1$, as we have already
noted. If stars in a globular cluster have a constant mass fraction of metals,
a He-rich star will appear to be more metal-rich than a He-normal star. The
positive correlations we find between $\Delta^{\rm X}$ and $\Delta^{\rm Na}$
are consistent with a He abundance variation since a Na-rich star is expected
to be He-rich relative to a Na-poor star. 

\citet{bragaglia10} examined a large sample of RGB stars in globular clusters
and argued that in addition to differences in metallicity, He-rich stars will
have subtly different temperatures and RGB bump luminosites. They found
evidence for all three effects in their sample. For their primordial (P) and
extreme (E) populations\footnote{A given star is assigned to a particular
population based on location in the [O/Fe] vs.\ [Na/Fe] plane according to
\citet{carretta09pie}.}, they found [Fe/H]$_{\rm E}$ $-$ [Fe/H]$_{\rm P}$ =
0.027 $\pm$ 0.010. The \citet{milone13} populations $a$ and $c$ may be regarded
as being equivalent to the \citet{carretta09pie} P and E populations,
respectively, and for the RGB bump sample we find $<\Delta^{\rm Fe}_{c}>$ $-$
$<\Delta^{\rm Fe}_{a}>$ = 0.039 $\pm$ 0.015, a value comparable to that of
\citet{bragaglia10}. If we consider all elements, the mean value $<\Delta^{\rm
X}_{c}>$ $-$ $<\Delta^{\rm X}_{a}>$ is 0.052 $\pm$ 0.005 ($\sigma$ = 0.019);
the smallest difference is for \crii\ (0.031 $\pm$ 0.023) and the largest
difference is for Si (0.092 $\pm$ 0.018). 

For a fixed value of $Z$, a change in helium abundance from $Y$=0.25 to
$Y$=0.28 would produce a change in [X/H] of +0.018 dex. By combining our
measurement errors with the expected 0.018 dex abundance variation due to He,
we can predict the abundance variations in [X/H]. If we compare these values
for each element to the observed variations, we find that the abundance
dispersions are, on average, 60\% $\pm$ 20\% larger than those expected from a
change in helium abundance of $\Delta Y$ = 0.03 combined with the measurement
uncertainties. Therefore, we tentatively conclude that while the observed
abundance variations are qualitatively consistent with a He variation, the
magnitudes of the observed variations are unlikely to be explained solely by a
He change of $\Delta Y$ = 0.03\footnote{The referee has pointed out that an
analysis of the colours and magnitudes of HB stars suggest a value of $\Delta
Y$ = 0.059 for NGC 6752 \citep{gratton10}. For such a value, He alone could
explain the abundance variations we find.}. To attribute the observed abundance
variations entirely to He would require $\Delta Y$ $\simeq$ 0.065, although
inclusion of 3D and/or NLTE effects could produce changes in the derived
differential abundances. Given the constraints on $\Delta Y$ from photometry
\citep{milone13}, some process in addition to the He variation may be required
to explain the abundance variations that we find. 

Before we consider another possibility, we briefly examine the data using
ATLAS12 model atmospheres \citep{castelli05,kurucz05,sbordone05}. We
constructed model atmospheres with \teff\ = 4800K, \logg\ = 2.0, \vt\ = 2.00
but with two different helium abundances $Y$ = 0.25 and $Y$ = 0.28. We also
ensured that the two models had the same mass fraction of metals, $Z$, and thus
they have slightly different metallicities $\Delta$[m/H] $\simeq$ 0.015. Using
these two model atmospheres, we computed abundances for all elements in three
RGB bump stars (9, 10 and 11).  These three stars have very similar stellar
parameters to the ATLAS12 models but they span a substantial range in Na
abundance. For a given element in a given star, we measured the abundance
difference when using the $Y$ = 0.28 vs.\ $Y$ = 0.25 models. The differences
are very small and essentially identical for all three stars; the average
abundance difference ($Y$ = 0.28 minus $Y$ = 0.25) is 0.001 dex $\pm$ 0.001 dex
($\sigma$ = 0.005 dex). That is, we obtain identical $Z$/$X$ ratios even though
the two models have different compositions. Such a result is expected given
that the line strength depends only on the ratio of the line opacity to
continuous opacity (H- for the program stars), i.e., the $Z$/$X$ ratio.  

\subsubsection{Inhomogeneous chemical evolution}

In the fourth scenario, we assume that the abundance variations are due to
chemical inhomogeneities in the pre- or proto-cluster environment. We
concentrate on the high statistical significance of the correlations between
($i$) Si and Na, ($ii$) Y and Na and ($iii$) Ca and Na seen in Figures
\ref{fig:allna}, \ref{fig:allnat}, \ref{fig:allna2} and \ref{fig:allnat2}.
Such correlations potentially provide great new insight into the origin of the
Na abundance variations in NGC 6752, and perhaps in all globular
clusters\footnote{In NGC 6397, \citet{lind11} found evidence for a possible
spread in yttrium abundance, 0.04 dex. In M4, \citet{villanova11} also found
evidence for a spread in yttrium abundance at the $\sim$0.1 dex level, although
\citet{dorazi13} do not confirm that result.}. 

The correlation between Si and Na could be attributed to leakage from the Mg-Al
chain into $^{28}$Si via $^{27}$Al(p,$\gamma$)$^{28}$Si during hydrogen burning
at high temperature \citep{ventura11}.  As noted already, similar conclusions
were drawn based on the correlations between Si and Al \citep{yong05} and Si
and N \citep{yong08nh}. To our knowledge, such correlations could arise from
both the asymptotic giant branch stars (AGB) and the fast rotating massive
stars (FRMS) scenarios. 

The correlation between Y and Na would suggest that the nucleosynthetic site
that produced Na also operated neutron-capture nucleosynthesis. To further
explore this issue, we derived chemical abundances for a larger suite of
elements expected to participate in neutron-capture reactions (Zn, Y, Zr, Ba,
La, Ce, Pr, Nd, Sm, Eu, and Dy).  We used only a subset of 10 RGB tip stars
with favorable stellar parameters (4250~$\leq T_{\rm eff} \leq$~4520~K; stars
mg8 to mg25) and followed the same procedure described in Sections 2.5 and 2.6,
using spectrum synthesis for all lines.  (The reference star was NGC 6752-mg9.)
For elements with only one measured line (Zn, Zr, Ba, Eu, and Dy), we adopted
0.02 dex as the ``fitting error'' and used this value as $\sigma_{\rm rand}$ in
the error analysis. For comparison, in our analysis of the 5380\AA\ La line in
Section 2.5, the average fitting error for the same 10 stars (mg8 to mg25) was
0.016 dex ($\sigma$ = 0.001 dex), and the minimum and maximum values were 0.015
and 0.017 dex, respectively. For the 6645\AA\ Eu line, the average fitting
error and minimum and maximum values were 0.017 dex ($\sigma$ = 0.001), 0.014
dex, and 0.018 dex, respectively. Therefore, we regard our choice of 0.02 dex
as a somewhat conservative estimate of the fitting error.  The line list and
abundance differences are presented in Tables \ref{tab:ncapline} and  
\ref{tab:abunncapb}.  With the exception of Sm, the average errors are
comparable to, or smaller than, the measured abundance dispersions. As before,
we take this as evidence for a genuine abundance dispersion, of small
amplitude, for these elements. 

\begin{table*}
 \centering
 \begin{minipage}{160mm}
  \caption{Line List for the Neutron-Capture Elements.\label{tab:ncapline}}
  \begin{tabular}{@{}cccrr@{}}
  \hline
         Wavelength & 
         Species\footnote{The digits to the left of the decimal point are the atomic
number. The digit to the right of the decimal point is the ionization state
(``0'' = neutral, ``1'' = singly ionised).} & 
         L.E.P & 
         $\log gf$ & 
         Source\footnote{1 = \citet{biemont11}; 
2 = \citet{denhartog03}; 
3 = \citet{ivarsson01}, using HFS from \citet{sneden09}; 
4 = \citet{lawler01la}; 
5 = \citet{lawler01la}, using HFS from \citet{ivans06}; 
6 = \citet{lawler01eu}, using HFS and isotope shifts from \citet{ivans06}; 
7 = \citet{lawler06}; 
8 = \citet{lawler09}; 
9 = \citet{li07}, using HFS from \citet{sneden09}; 
10 = \citet{ljung06}; 
11 = \citet{fuhr09}; 
12 = \citet{roederer12}; 
13 = \citet{wickliffe00}}\\
         \AA & 
          & 
         eV & 
          & 
          \\ 
         (1) & 
         (2) &
         (3) &
         (4) & 
         (5) \\
\hline 
 4810.53 &  30.0 &  4.08 &  $-$0.15 &  12 \\
 4883.68 &  39.1 &  1.08 &     0.19 &   1 \\
 4900.12 &  39.1 &  1.03 &     0.03 &   1 \\
 4982.13 &  39.1 &  1.03 &  $-$1.32 &   1 \\
 5087.42 &  39.1 &  1.08 &  $-$0.16 &   1 \\
 5119.11 &  39.1 &  0.99 &  $-$1.33 &   1 \\
 5205.72 &  39.1 &  1.03 &  $-$0.28 &   1 \\
 5289.82 &  39.1 &  1.03 &  $-$1.68 &   1 \\
 5402.77 &  39.1 &  1.84 &  $-$0.31 &   1 \\
 5473.38 &  39.1 &  1.74 &  $-$0.78 &   1 \\
 5544.61 &  39.1 &  1.74 &  $-$0.83 &   1 \\
 5728.89 &  39.1 &  1.84 &  $-$1.15 &   1 \\
 5112.27 &  40.1 &  1.66 &  $-$0.85 &  10 \\
 6496.90 &  56.1 &  0.60 &  $-$0.41 &  11 \\
 5114.56 &  57.1 &  0.24 &  $-$1.03 &   5 \\
 5122.99 &  57.1 &  0.32 &  $-$0.91 &   5 \\
 5290.82 &  57.1 &  0.00 &  $-$1.65 &   4 \\
 5301.97 &  57.1 &  0.40 &  $-$0.94 &   5 \\
 5303.53 &  57.1 &  0.32 &  $-$1.35 &   5 \\
 5482.27 &  57.1 &  0.00 &  $-$2.23 &   5 \\
 6262.29 &  57.1 &  0.40 &  $-$1.22 &   5 \\
 6390.48 &  57.1 &  0.32 &  $-$1.41 &   5 \\
 5274.23 &  58.1 &  1.04 &     0.13 &   8 \\
 5330.56 &  58.1 &  0.87 &  $-$0.40 &   8 \\
 6043.37 &  58.1 &  1.20 &  $-$0.48 &   8 \\
 5259.73 &  59.1 &  0.63 &     0.11 &   3 \\
 5322.77 &  59.1 &  0.48 &  $-$0.12 &   9 \\
 4797.15 &  60.1 &  0.56 &  $-$0.69 &   2 \\
 4825.48 &  60.1 &  0.18 &  $-$0.42 &   2 \\
 4914.38 &  60.1 &  0.38 &  $-$0.70 &   2 \\
 4959.12 &  60.1 &  0.06 &  $-$0.80 &   2 \\
 4987.16 &  60.1 &  0.74 &  $-$0.79 &   2 \\
 5063.72 &  60.1 &  0.98 &  $-$0.62 &   2 \\
 5092.79 &  60.1 &  0.38 &  $-$0.61 &   2 \\
 5130.59 &  60.1 &  1.30 &     0.45 &   2 \\
 5132.33 &  60.1 &  0.56 &  $-$0.71 &   2 \\
 5234.19 &  60.1 &  0.55 &  $-$0.51 &   2 \\
 5249.58 &  60.1 &  0.98 &     0.20 &   2 \\
 5293.16 &  60.1 &  0.82 &     0.10 &   2 \\
 5306.46 &  60.1 &  0.86 &  $-$0.97 &   2 \\
 5311.45 &  60.1 &  0.98 &  $-$0.42 &   2 \\
 5319.81 &  60.1 &  0.55 &  $-$0.14 &   2 \\
 5356.97 &  60.1 &  1.26 &  $-$0.28 &   2 \\
 5485.70 &  60.1 &  1.26 &  $-$0.12 &   2 \\
 4815.81 &  62.1 &  0.18 &  $-$0.82 &   7 \\
 4844.21 &  62.1 &  0.28 &  $-$0.89 &   7 \\
 4854.37 &  62.1 &  0.38 &  $-$1.25 &   7 \\
 4913.26 &  62.1 &  0.66 &  $-$0.93 &   7 \\
 6645.10 &  63.1 &  1.38 &     0.12 &   6 \\
 5169.69 &  66.1 &  0.10 &  $-$1.95 &  13 \\
\hline
\end{tabular}
\end{minipage}
\end{table*}

\begin{table*}
 \centering
 \begin{minipage}{160mm}
  \caption{Differential Abundances For Neutron-Capture Elements (Zn, Y, Zr, Ba,
La and Ce) in a Subset of RGB Tip Stars (Reference Star =  NGC
6752-mg9).\label{tab:abunncapb}} 
  \begin{tabular}{@{}lrcrcrcrcrcrcrcrc@{}}
  \hline
         Star & 
         $\Delta^{\rm Zn}$ &
         $\sigma$ & 
         $\Delta^{\rm Y}$ &
         $\sigma$ & 
         $\Delta^{\rm Zr}$ &
         $\sigma$ & 
         $\Delta^{\rm Ba}$ &
         $\sigma$ & 
         $\Delta^{\rm La}$ &
         $\sigma$ & 
         $\Delta^{\rm Ce}$ &
         $\sigma$ \\ 
\hline 
NGC6752-mg8  & 	$-$0.080 & 0.025 &    0.020 & 0.017 &    0.040 & 0.021 & $-$0.130 & 0.035 &    0.025 & 0.016 &    0.007 & 0.024 \\ 
NGC6752-mg10 & 	$-$0.080 & 0.025 &    0.075 & 0.016 &    0.070 & 0.024 &    0.010 & 0.039 &    0.022 & 0.016 &    0.017 & 0.030 \\ 
NGC6752-mg12 & 	$-$0.060 & 0.029 & $-$0.042 & 0.020 &    0.020 & 0.023 & $-$0.030 & 0.036 & $-$0.008 & 0.026 &    0.043 & 0.033 \\ 
NGC6752-mg15 & 	$-$0.050 & 0.033 & $-$0.014 & 0.023 &    0.000 & 0.024 & $-$0.090 & 0.045 &    0.003 & 0.011 &    0.037 & 0.035 \\ 
NGC6752-mg18 & 	$-$0.050 & 0.033 &    0.021 & 0.023 &    0.020 & 0.024 & $-$0.070 & 0.044 &    0.029 & 0.017 &    0.047 & 0.033 \\ 
NGC6752-mg21 & 	   0.010 & 0.031 &    0.060 & 0.020 &    0.040 & 0.024 & $-$0.020 & 0.043 &    0.045 & 0.019 &    0.007 & 0.038 \\ 
NGC6752-mg22 & 	$-$0.020 & 0.026 &    0.045 & 0.019 &    0.080 & 0.021 & $-$0.020 & 0.062 &    0.039 & 0.013 &    0.077 & 0.020 \\ 
NGC6752-mg24 & 	$-$0.050 & 0.037 & $-$0.082 & 0.031 & $-$0.020 & 0.025 & $-$0.150 & 0.065 & $-$0.004 & 0.012 & $-$0.053 & 0.030 \\ 
NGC6752-mg25 & 	$-$0.130 & 0.025 & $-$0.016 & 0.024 &    0.040 & 0.020 & $-$0.150 & 0.038 & $-$0.025 & 0.010 & $-$0.010 & 0.025 \\ 
\hline
\end{tabular}
\end{minipage}
\\
In order to place the above values onto an absolute scale, the absolute
abundances we obtain for the reference stars are given below. We caution,
however, that the absolute scale has not been critically evaluated (see Section
2.5 for more details). 
\\ 
NGC6752-mg9: 
A(Zn) = 3.02, 
A(Y)  = 0.49, 
A(Zr) = 1.34, 
A(Ba) = 1.02, 
A(La) = $-$0.33, 
A(Ce) = 0.00, 
\end{table*}

\begin{table*}
 \centering
 \begin{minipage}{160mm}
  \caption{Differential Abundances For Neutron-Capture Elements (Pr, Nd, Sm, Eu
and Dy) in a Subset of RGB Tip Stars (Reference Star =  NGC
6752-mg9).\label{tab:abunncapa}} 
  \begin{tabular}{@{}lrcrcrcrcrcrcrc@{}}
  \hline
         Star & 
         $\Delta^{\rm Pr}$ &
         $\sigma$ & 
         $\Delta^{\rm Nd}$ &
         $\sigma$ & 
         $\Delta^{\rm Sm}$ &
         $\sigma$ & 
         $\Delta^{\rm Eu}$ &
         $\sigma$ & 
         $\Delta^{\rm Dy}$ &
         $\sigma$ \\ 
\hline 
NGC6752-mg8  &    0.010 & 0.011 &    0.038 & 0.016 & $-$0.037 & 0.029 &    0.070 & 0.021 &    0.040 & 0.021 \\ 
NGC6752-mg10 &    0.005 & 0.021 &    0.029 & 0.016 & $-$0.048 & 0.029 &    0.080 & 0.023 &    0.110 & 0.021 \\ 
NGC6752-mg12 &    0.010 & 0.040 &    0.025 & 0.012 & $-$0.022 & 0.016 &    0.030 & 0.022 &    0.040 & 0.021 \\ 
NGC6752-mg15 & $-$0.040 & 0.016 &    0.034 & 0.011 & $-$0.010 & 0.027 &    0.040 & 0.025 &    0.130 & 0.021 \\ 
NGC6752-mg18 & $-$0.005 & 0.025 &    0.019 & 0.017 & $-$0.030 & 0.025 &    0.020 & 0.024 &    0.070 & 0.021 \\ 
NGC6752-mg21 &    0.005 & 0.035 &    0.057 & 0.014 &    0.005 & 0.047 &    0.050 & 0.024 &    0.100 & 0.021 \\ 
NGC6752-mg22 &    0.005 & 0.022 &    0.036 & 0.017 &    0.010 & 0.024 &    0.030 & 0.021 &    0.140 & 0.022 \\ 
NGC6752-mg24 & $-$0.030 & 0.010 & $-$0.012 & 0.017 & $-$0.030 & 0.026 &    0.000 & 0.027 &    0.090 & 0.021 \\ 
NGC6752-mg25 & $-$0.045 & 0.031 & $-$0.008 & 0.017 & $-$0.065 & 0.045 &    0.000 & 0.021 &    0.020 & 0.021 \\ 
\hline
\end{tabular}
\end{minipage}
\\
In order to place the above values onto an absolute scale, the absolute
abundances we obtain for the reference stars are given below. We caution,
however, that the absolute scale has not been critically evaluated (see Section
2.5 for more details). 
\\ 
NGC6752-mg9: 
A(Pr) = $-$0.75, 
A(Nd) = $-$0.02, 
A(Sm) = $-$0.38, 
A(Eu) = $-$0.69, 
A(Dy) = $-$0.25
\end{table*}

\begin{figure}
\centering
      \includegraphics[width=0.9\hsize]{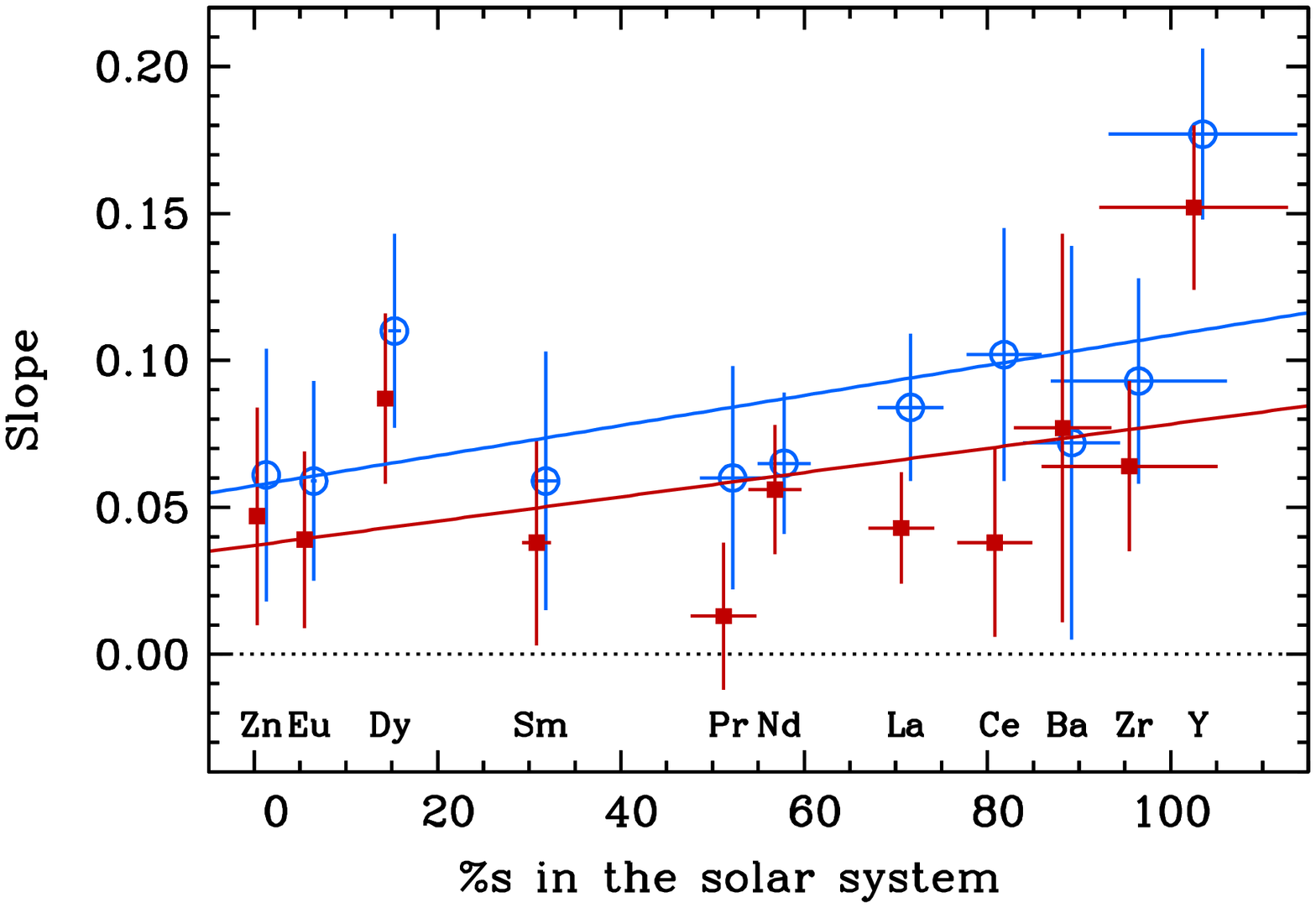} 
      \caption{Slope of the fit to $\Delta^{\rm X}$ vs.\ $\Delta^{\rm Na}$ vs.\
percentage attributed to the $s$-process in solar system material (using the
\citealt{bisterzo11} values). The red squares are from the ``regular'' analysis
while the blue open circles are fits to the data when abundance trends with
\teff\ have been removed, i.e., slopes of the fits to $\Delta^{\rm X}_{\rm T}$
vs.\ $\Delta^{\rm Na}_{\rm T}$.  Small horizontal offsets ($\pm$~0.5\%) have
been applied to aid visibility.  Neither slope is significant at the 2$\sigma$
level.}
         \label{fig:ncaprat}
\end{figure}

For these new measurements, we fit the slope to $\Delta^{\rm X}$ vs.\
$\Delta^{\rm Na}$ as in Section 3.2. We find that the slope is positive for all
elements. If we remove the abundance trends with \teff\ as described in Section
3.4, these results remain unchanged. For Y, La, Nd, and Eu, the results from
this new analysis are in agreement with the previous results (at the
$<$3$\sigma$ level).  In Figure \ref{fig:ncaprat}, we plot the slope of the
fit to $\Delta^{\rm X}$ vs.\ $\Delta^{\rm Na}$ against the percentage
attributed to the $s$-process in the solar system, adopting the solar
$s$-process percentages calculated by \citet{bisterzo11}.  In this figure, we
also show the slopes when fitting $\Delta^{\rm X}_{\rm T}$ vs.\ $\Delta^{\rm
Na}_{\rm T}$, i.e., after removing the abundance trends with \teff. In both
cases, the slopes are not of high statistical significance, $<$2$\sigma$
level.  If we exclude Y, a possible outlier, the slopes are of even lower
statistical significance, $<$1$\sigma$.  (The neighboring elements Y and Zr
are both members of the first $s$-process peak, so we would not expect their
nucleosynthesis histories to be substantially different.) The absence of a
significant trend in Figure \ref{fig:ncaprat} suggests that the abundance
variations are not the result of preferentially introducing more $s$-process
material than $r$-process material. 

The correlation between Ca and Na requires massive stars to have played a role
in the pre- or proto-cluster environment since the synthesis of Ca is believed
to occur primarily during O-burning and Si-burning in those objects
\citep{clayton03}. That said, the abundances for all elements are positively
correlated with the Na abundance, and for any pair of elements heavier than Si,
the abundances are positively correlated. Furthermore, the ratios for any pair
of elements (e.g., $\Delta^{\rm Ni}$ $-$ $\Delta^{\rm Ca}$ using our
terminology) are constant at the 0.036 dex $\pm$ 0.001 dex ($\sigma$ = 0.012)
level for the RGB tip sample (excluding Eu, which has considerably larger
measurement errors) and essentially identical results are found for the RGB
bump sample.  Thus, the origin of such correlations demands a source (or
sources) capable of synthesis of Na, $\alpha$, Fe-peak and neutron-capture
elements and this diverse suite of elements must be synthesized in essentially
equal amounts.  No individual star can achieve such nucleosynthesis, and
therefore, a variety of sources is required. 

The underlying assumption in this work, and in other studies, is that the
star-to-star light element abundance variations in mono-metallic globular
clusters are produced by some source (AGB, FRMS and/or massive binaries)
within the duration of star formation in the globular cluster. Such an
assumption appears reasonable, although unresolved issues related to
nucleosynthesis and enrichment timescales remain (e.g.,
\citealt{fenner04,decressin07,prantzos07,pumo08,demink09,dercole12}).
Regarding the heavy elements, one might also assume that the star-to-star
abundance variations and correlations with Na are produced by some source
within the duration of star formation in this globular cluster, provided the
heavy elements are produced in the same ratios as those already found in the
first generation stars.  Another possibility is that the heavy element
abundance variations and correlations with Na arise because the ejecta from the
source that produced Na was diluted into gas with slightly higher [X/H] ratios
that entered the cluster while the later generations of stars formed.  In this
scenario, production of the light elements, including Na, is completely
decoupled from production of all elements heavier than Si.  Unfortunately,
there are no obvious observational tests to distinguish between these two
scenarios.  We thus regard the ``production during cluster formation'' and
``dilution with pristine material'' scenarios as equally valid possibilities
for the abundance variations. 

The penultimate issue we raise concerns whether the distribution of the heavy
element abundances is discrete or continuous.  As noted in Section 2.1,
\citet{milone13} have identified three stellar populations in NGC 6752 based on
\textit{HST} and ground-based Str{\" o}mgren photometry. The three populations
can be found at all evolutionary stages (main sequence, subgiant branch and red
giant branch). Additionally, each population exhibits distinct chemical
abundance patterns for the light elements (e.g., N, O, Na, Mg and Al). In
Figure \ref{fig:fena}, populations $a$ (green), $b$ (magenta) and $c$ (blue)
have distinct $\Delta^{\rm Na}$ abundances. In Figures \ref{fig:sica} and
\ref{fig:nind} (and other figures), we use the same colour scheme to denote the
three populations. In general, population $c$ (blue) exhibits a larger (i.e.,
more positive) value for $\Delta^{\rm X}$ than population $a$ (green), while
population $b$ (magenta) lies between populations $a$ and $c$.  Such a result
is expected given ($i$) the Na abundances of each population and ($ii$) the
correlation between $\Delta^{\rm X}$ and $\Delta^{\rm Na}$. Although we have
achieved very high precision relative abundance measurements, it is not clear
whether the abundance distributions seen in Figures \ref{fig:sica} and
\ref{fig:nind} are consistent with three discrete values in the $\Delta^{\rm
X}$ vs. $\Delta^{\rm Y}$ plane, corresponding to the \citet{milone13}
populations $a$, $b$ and $c$. (That said, it is not obvious whether the
\citealt{milone13} data show three discrete photometric sequences.) Additional
studies may be necessary to clarify whether the heavy element abundance
distribution is discrete or continuous in this globular cluster. 

Finally, we mentioned in the introduction that \citet{sneden05} examined the
[Ni/Fe] ratio in the context of cluster abundance accuracy limits. There was an
apparent limit in $\sigma$[Ni/Fe] at the $\sim$0.06 dex level. For the RGB tip
and RGB bump samples, we find $\sigma$($\Delta^{\rm Ni}$ $-$ $\Delta^{\rm Fe}$)
= 0.009 and 0.010, respectively, thereby highlighting the great improvement in
abundance precision that can be obtained when conducting a strictly
differential analysis of high quality spectra. 

\section{SUMMARY}

We have obtained very high precision chemical abundance measurements,
$\Delta^{\rm X}$, through a strictly differential analysis of high quality UVES
spectra of giant stars in the globular cluster NGC 6752. The measurement
uncertainties and average uncertainties for a given element,
$<\sigma\Delta^{\rm X}>$, are as low as $\sim$0.01 dex. The observed abundance
dispersions, and abundance dispersions about various linear fits (e.g.,
$\Delta^{\rm X}$ vs.\ \teff\ or $\Delta^{\rm X}$ vs.\ $\Delta^{\rm Y}$), are
often considerably larger than the average abundance uncertainty. We find
positive correlations between any given element and Na, i.e., $\Delta^{\rm X}$
vs.\ $\Delta^{\rm Na}$, and indeed for any combination of elements, e.g.,
$\Delta^{\rm X}$ vs.\ $\Delta^{\rm Y}$. These correlations are often of high
statistical significance ($>$ 5$\sigma$), although we note that the amplitudes
of the abundance variations are small.  These results are unchanged even after
removing abundance trends with \teff\ and/or when using a different pair of
reference stars. Indeed, the likelihood of these results being due to random
error is exceedingly small. Therefore, we argue that there is a genuine
abundance dispersion in this cluster, at the $\sim$0.03 dex level. 

In order to explain these results, we consider four possibilities. The
abundance variations and correlations may reflect ($i$) systematic errors in
the stellar parameters, ($ii$) star-to-star CNO abundance variations, ($iii$)
star-to-star He abundance variations and/or ($iv$) inhomogeneous chemical
evolution. In the context of point ($i$), the stellar parameter uncertainties
would require substantial increases; our results are seen for all elements
(covering a range of ionization potentials and ionization states) and no single
change in \teff, \logg\ or \vt\ would remove the abundance correlations for all
elements. Regarding point ($ii$), predictions by \citet{drake93} suggest that
for weak lines such as those in this study, using model atmospheres with
appropriate CNO abundances will not change our results.  Regarding point
($iii$), for a fixed mass fraction of metals ($Z$), an increase in helium
abundance ($Y$) would result in a lower hydrogen abundance ($X$) and therefore
a higher metal-to-hydrogen ratio, $Z$/$X$.  Since Na and He abundances are
expected to be correlated, the positive correlations we find between
$\Delta^{\rm X}$ vs.\ $\Delta^{\rm Na}$ are consistent with a He abundance
variation (for constant $Z$). Given the current constraints on $\Delta Y$ from
photometry \citep{milone13}, it is likely that the abundance variations cannot
be attributed solely to He. Nevertheless, He abundance variations probably play
an important role in producing the abundance variations that we find.
Concerning point ($iv$), the correlation between Si and Na could arise from
leakage from the Mg-Al chain into Si in either AGB or FRMS. For the
neutron-capture elements, there is no significant trend between the slope of
the fit to $\Delta^{\rm X}$ vs.\ $\Delta^{\rm Na}$ when plotted against
percentage attributed to the $s$-process in solar system material.  Thus, their
abundance variations are probably not related to $s$-process production by
whatever source produced the light element variations.  That all elements are
correlated requires a nucleosynthetic source(s) capable of synthesizing Na,
$\alpha$, Fe-peak and neutron-capture elements.  Additionally,
element-to-element ratios (e.g., $\Delta^{\rm Ni}$ $-$ $\Delta^{\rm Ca}$ using
our terminology) are constant at the $\sim$0.03 dex level. No individual object
can achieve the required nucleosynthesis.  We cannot ascertain whether the
heavy elements were produced ($a$) within the duration of star formation in
this globular cluster or ($b$) by dilution of Na-rich material into gas with
slightly higher [X/H] ratios that entered the cluster while the second (and
later) generations of stars formed.  In summary, our results may be explained
by some combination of He abundance variations and inhomogeneous chemical
evolution (i.e., metallicity variations). There may be other explanations for
the observed abundance variations and correlations.  Nevertheless, we encourage
similar studies of other globular clusters with no obvious dispersion in
Fe-peak elements. 

\section*{Acknowledgments}

We warmly thank the referee, Raffaele Gratton, for helpful comments that
improved and clarified this work. We thank J.\ A.\ Johnson and A.\ I.\ Karakas
for helpful discussions.  D.\ Y., J.\ E.\ N., A.\ P.\ M., A.\ F.\ M., R.\ C.\
and M.\ A.\ gratefully acknowledge support from the Australian Research Council
(grants DP0984924, FL110100012, DP120100475, DP120100991 and DE120102940). J.\
M.\ would like to acknowledge support from FAPESP (2010/17510-3; 2012/24392-2)
and CNPq (Bolsa de Produtividade).  Funding for the Stellar Astrophysics Centre
is provided by The Danish National Research Foundation. The research is
supported by the ASTERISK project (ASTERoseismic Investigations with SONG and
Kepler) funded by the European Research Council (Grant agreement no.: 267864).
I.~U.~R.\ is grateful for support from the Carnegie Institution for Science
through the Barbara McClintock Fellowship.  P.~C.\ acknowledges support from
FAPESP Project 2008/58406-4.

\label{lastpage}

\end{document}